\documentclass[twocolumn,trackchanges,twocolappendix]{aastex7}
\usepackage{amsmath}
\usepackage{graphicx}
\usepackage{soul}
\usepackage{hyperref}
\usepackage{multirow}

\newcommand{\ramsesrt}{\textsc{ramses-rt}}
\newcommand{\ramses}{\textsc{ramses}}
\newcommand{\pgalf}{\texttt{PGalF}}
\newcommand{\darwin}{\texttt{DARWIN}}
\newcommand{\darwini}{\texttt{DARWIN-1}}
\newcommand{\darwinii}{\texttt{DARWIN-2}}
\newcommand{\darwiniii}{\texttt{DARWIN-3}}
\newcommand{\skirt}{\textsc{skirt}}
\newcommand{\powderday}{\textsc{powderday}}

\definecolor{red}{RGB}{250,0,0}
\definecolor{magenta}{RGB}{167,0,135}

\definecolor{pink}{RGB}{255,51,153}

\definecolor{green}{RGB}{0,153,0}

\definecolor{blue}{RGB}{0,0,250}

\definecolor{grey}{RGB}{180,180,180}

\setstcolor{red}

\usepackage{soul} 
\usepackage{xcolor}
\definecolor{aquamarine}{rgb}{0.5, 1.0, 0.83}
\definecolor{cyan}{rgb}{0.8784,1.0,1.0}

\newcommand{\msun}{\mbox{$M_{\odot}$}}

\newcommand{\kms}{\mbox{${\rm km\,s^{-1}}$}}

\newcommand{\nH}{\mbox{$n_{\rm H}$}}

\newcommand{\mhalo}{\mbox{$M_{\rm halo}$}}
\newcommand{\cmq}{\mbox{$\,{\rm cm^{-3}}$}}

\received{\today}
\revised{\today}

\shorttitle{DARWIN project}
\shortauthors{DARWIN collaborators}

\begin{document}

\title{The DARWIN Project: Cosmological Simulations of Galaxy Formation down to Dwarf Scales in Diverse Environments}

\author[0000-0002-6810-1778]{Jaehyun Lee}
\affiliation{Korea Astronomy and Space Science Institute, 776, Daedeokdae-ro, Yuseong-gu, Daejeon 34055, Republic of Korea}
\email[show]{jaehyun@kasi.re.kr}

\author[0000-0001-9544-7021]{Kyungwon Chun}
\affiliation{Korea Astronomy and Space Science Institute, 776, Daedeokdae-ro, Yuseong-gu, Daejeon 34055, Republic of Korea}\email{kwchun@kasi.re.kr}

\author[0000-0002-3950-3997]{Taysun Kimm}
\affiliation{Department of Astronomy, Yonsei University, 50 Yonsei-ro, Seodaemun-gu, Seoul 03722, Republic of Korea}\email{tkimm@yonsei.ac.kr}

\author[0000-0003-4164-5414]{Yonghwi Kim}
\affiliation{Korea Institute of Science and Technology Information (KISTI), 245 Daehak-ro, Yuseong-gu, Daejeon 34141, Republic of Korea}\email{yonghwi.kim@kisti.re.kr}

\author[0000-0001-5135-1693]{Jihye Shin}
\affiliation{Korea Astronomy and Space Science Institute, 776, Daedeokdae-ro, Yuseong-gu, Daejeon 34055, Republic of Korea}\affiliation{Department of Astronomy and Space Science, University of Science and Technology, Daejeon 34113, Republic of Korea}\email[show]{jhshin@kasi.re.kr}

\author[0000-0002-8131-6378]{Ena Choi}
\affiliation{Department of Physics, University of Seoul, 163 Seoulsiripdaero, Dongdaemun-gu, Seoul 02504, Republic of Korea}\email{enachoi@uos.ac.kr}

\author[0000-0003-4923-8485]{Sungwook E. Hong}
\affiliation{Korea Astronomy and Space Science Institute, 776, Daedeokdae-ro, Yuseong-gu, Daejeon 34055, Republic of Korea}\affiliation{Department of Astronomy and Space Science, University of Science and Technology, Daejeon 34113, Republic of Korea}\email{swhong@kasi.re.kr}

\author[0000-0001-6529-9777]{Myoungwon Jeon}
\affiliation{School of Space Research, Kyung Hee University, 1732 Deogyeong-daero, Giheung-gu, Yongin-si, Gyeonggi-do 17104, Republic of Korea}
\affiliation{Department of Astronomy \& Space Science, Kyung Hee University, 1732 Deogyeong-daero, Yongin-si, Gyeonggi-do 17104, Republic of Korea}
\email{myjeon@khu.ac.kr}

\author[0000-0001-6470-7476]{Hannah Jhee}
\affiliation{Department of Physics, University of Seoul, 163 Seoulsiripdaero, Dongdaemun-gu, Seoul 02504, Republic of Korea}\email{hannah.jhee@uos.ac.kr}

\author[0000-0002-9144-1383]{Minyong Jung}
\affiliation{Center for Theoretical Physics, Department of Physics and Astronomy, Seoul National University, Seoul 08826, Republic of Korea}
\email{wispedia@snu.ac.kr}

\author[0000-0002-7820-2281]{Hyeonyong Kim}
\affiliation{Center for Theoretical Physics, Department of Physics and Astronomy, Seoul National University, Seoul 08826, Republic of Korea}\email{gusdyd398@snu.ac.kr}

\author[0000-0003-4464-1160]{Ji-hoon Kim}
\affiliation{Center for Theoretical Physics, Department of Physics and Astronomy, Seoul National University, Seoul 08826, Republic of Korea}
\affiliation{Seoul National University Astronomy Research Center, Seoul 08826, Republic of Korea}
\affiliation{Institute for Data Innovation in Science, Seoul National University, Seoul 08826, Republic of Korea}\email{mornkr@snu.ac.kr}

\author[0000-0002-4391-2275]{Juhan Kim}
\affiliation{Center for Advanced Computation, Korea Institute for Advanced Study, 85 Hoegiro, Dongdaemun-gu, Seoul 02455, Republic of Korea}\email{kjhan@kias.re.kr}

\author{Oh-Kyoung Kwon}
\affiliation{Korea Institute of Science and Technology Information (KISTI), 245 Daehak-ro, Yuseong-gu, Daejeon 34141, Republic of Korea}\email{okkwon@kisti.re.kr}

\author[0000-0002-4362-4070]{Hyunmi Song}
\affiliation{Department of Physics, University of Seoul, 163 Seoulsiripdaero, Dongdaemun-gu, Seoul 02504, Republic of Korea}\email{hmsong@cnu.ac.kr}

\author[0000-0003-0225-6387]{Yohan Dubois}
\affiliation{Institut d’Astrophysique de Paris, UMR 7095, CNRS, Sorbonne Université, 98 bis boulevard Arago, F-75014 Paris, France}
\email{dubois@iap.fr}

\author[orcid=0000-0001-7670-4897,sname='Ha']{Ji-Hoon Ha}
\affiliation{Korea Astronomy and Space Science Institute, 776, Daedeokdae-ro, Yuseong-gu, Daejeon 34055, Republic of Korea}\email{jhha@kasi.re.kr}

\author[0000-0002-2624-3129]{Daniel Han}
\affiliation{Department of Astronomy, Yonsei University, 50 Yonsei-ro, Seodaemun-gu, Seoul 03722, Republic of Korea}\email{daniel.han@yonsei.ac.kr}

\author[0000-0001-9939-713X]{San Han}\email{san.han@iap.fr}
\affil{Institut d’Astrophysique de Paris, UMR 7095, CNRS, Sorbonne Université, 98 bis boulevard Arago, F-75014 Paris, France}

\author[0009-0005-5530-9958]{Jun Yong Park}
\affiliation{Department of Physics and Astronomy, Seoul National University, Seoul 08826, Republic of Korea}\email{tldns4755@snu.ac.kr}

\author[0000-0003-0695-6735]{Christophe Pichon}
\affiliation{Institut d’Astrophysique de Paris, UMR 7095, CNRS, Sorbonne Université, 98 bis boulevard Arago, F-75014 Paris, France}
\affiliation{Department of Astronomy \& Space Science, Kyung Hee University, 1732 Deogyeong-daero, Yongin-si, Gyeonggi-do 17104, Republic of Korea}
\email{pichon@iap.fr}

\begin{abstract}
We present the \darwin\ project, a suite of cosmological simulations with \ramses\ to study dwarf-galaxy formation and evolution across diverse environments. \darwini\ is a full-box simulation in a cubic volume of $(65.5\,{\rm cMpc})^3$ with a minimum cell size of $500$~pc, while \darwinii\ comprises zoom-in simulations of 24 field regions and a Milky Way analog connected to a filament, reaching $62.5$~pc resolution. The simulations employ baryonic physics tailored to their aims. For \darwinii, we include radiative transfer in three photon energy bands to trace hydrogen and helium ionization. \darwini\ reproduces the cosmic star formation rate density and galaxy stellar mass functions reasonably well at $z>1$. Both simulations broadly reproduce observed stellar mass--halo mass relations, chemical abundances, star-formation surface densities, and black hole--galaxy scaling relations. At lower redshift, \darwini\ shows an excess in the number density of massive galaxies, although the inferred stellar content of massive halos depends on the definition of galaxy stellar mass. Other systematic differences arise from spatial resolution, feedback efficiency, and environmental selection. To facilitate comparison with observations, we generate mock images using the radiative transfer codes \skirt\ and \powderday. We also explore a machine-learning approach to mitigate resolution-driven biases in the dwarf galaxy properties of \darwini\ using the higher-resolution \darwinii\ simulations. The \darwin\ simulation suite therefore provides a comprehensive numerical framework for studying the formation, evolution, and internal physics of dwarf galaxies in a cosmological context.
\end{abstract}


\section{Introduction}\label{intro} 

Dwarf galaxies are among the faintest and least massive galaxies, yet they likely constitute the most abundant galaxy population in the Universe \citep{mateo98,tolstoy09,mcconnachie12}. Within the $\Lambda$ cold dark matter ($\Lambda$CDM) paradigm, massive galaxies form through the hierarchical assembly of smaller systems~\citep{white78,blumenthal84}, making dwarfs central to our understanding of galaxy formation~\citep{bullock17}. Moreover, dwarf galaxies that survive to the present day may retain imprints of the early stages of galaxy formation \citep{ricotti05,frebel15}, providing a unique window into the conditions under which the first generations of galaxies formed and evolved \citep{tolstoy09,bullock17}. 

Compared with more massive systems, dwarf galaxies exhibit substantially larger scatter in several fundamental scaling relations, including the mass-size relation \citep{brasseur11,danieli18}, mass-metallicity relation \citep{kirby13}, and stellar mass-halo mass relation \citep{behroozi13,moster13,garrison17}. Recent deep observations have further revealed considerable diversity among dwarf galaxies, including ultra-diffuse galaxies, ultra-faint dwarf galaxies, and quenched isolated dwarfs \citep{vandokkum15,simon19,carlsten22}. Part of this diversity may arise from their shallow potential wells, which allow processes such as cosmic reionization and supernova feedback to strongly regulate their gas content and star formation histories~\citep{dekel86,efstathiou92,bullock00,governato10,hopkins14,wheeler15}, while also making dwarf galaxies more susceptible to environmental effects \citep{mayer06,slater14,simpson18}. The observed diversity therefore provides a sensitive probe of the physical processes that regulate low-mass galaxy evolution \citep{weisz14,bullock17}. 

Despite its success in reproducing the large-scale structure of the Universe \citep[e.g.,][]{springel05}, the $\Lambda$CDM model faces several long-standing challenges on dwarf-galaxy scales \citep{bullock17,Weinberg2015}. These include the missing-satellites problem~\citep{Klypin1999,Moore1999}, the cusp-core problem~\citep{Flores1994,Moore1994,deBlok2010}, and the too-big-to-fail problem \citep{BoylanKolchin2011,BoylanKolchin2012}, which have motivated extensive observational and numerical investigations within the $\Lambda$CDM framework~\citep{Governato2012,bullock17}. Although dwarf galaxies are abundant across diverse environments \citep{mcconnachie12,carlsten22}, their faintness makes them difficult to detect \citep{simon19}. Some of these apparent discrepancies may therefore arise, at least in part, from observational incompleteness~\citep[e.g.,][]{koposov08,newton18,kim23}.

Numerical modeling of dwarf galaxies is also challenging because high spatial and mass resolution is required to accurately resolve their internal structure~\citep{Hopkins2018,wheeler19, Andersson2025}. Their shallow gravitational potentials also make their evolution particularly sensitive to the treatment of baryonic physics \citep{Pontzen2012,Read2016,rey25}. Environmental processes further shape dwarf galaxy evolution \citep{mayer06,Emerick2016}, limiting the ability of isolated simulations to capture the full diversity of the population \citep{briggs26}. Capturing the environmental diversity of dwarf galaxies while resolving their internal structure would ideally require representative cosmological volumes together with resolution on $\sim100\,$pc scales, a combination that remains computationally demanding~\citep{Vogelsberger2020,kannan22}.

Cosmological zoom-in simulations offer a practical way to alleviate these difficulties \citep{Katz1993,Onorbe2014}. By applying high resolution only to selected regions while using lower resolution elsewhere, zoom-in simulations can follow structure formation in detail while retaining the long-range gravitational influence of the surrounding large-scale structure at substantially lower computational cost~\citep[e.g.,][]{Kim2014,dubois21,lee21,han26}. More broadly, cosmological simulations of dwarf galaxies have pursued several complementary strategies: 1) simulation suites designed to provide statistically meaningful samples across a broad range of dwarf-galaxy masses (\texttt{NIHAO}; \citealt{wang15}, \texttt{MARVEL-ous dwarfs}; \citealt{Bellovary19}, \texttt{TNG50}; \citealt{Pillepich19}), 2) simulations targeting specific environments, such as Milky Way-like, Local Group-like or cluster regions (\texttt{Auriga}; \citealt{grand17}, \texttt{APOSTLE}; \citealt{sawala16}, \texttt{Romulus}; \citealt{Tremmel2017MNRAS}, \texttt{DC Justice League}; \citealt{Applebaum2021}, \texttt{NewHorizon} \citealt{dubois21}, \texttt{FIREbox}; \citealt{feldmann23}, \texttt{NewCluster}; \citealt{han26}), and 3) ultra-high resolution simulations that incorporate detailed baryonic physics and feedback processes (\texttt{EDGE}; \citealt{rey25}, \texttt{FIRE-2/Latte}; \citealt{wetzel16, Hopkins2018}, \texttt{RIGEL}; \citealt{Deng2024}, \texttt{AEOS}; \citealt{Brauer2025}). Although these simulations have significantly improved our understanding of dwarf galaxies, they remain limited in their ability to simultaneously capture diverse environments and resolve detailed internal structures within a consistent numerical framework. 

To address this limitation, we introduce \darwin\ (DAzzling Realization of dWarf galaxies In the Next generation of cosmological hydrodynamic simulations), a hierarchical cosmological simulation framework that combines three nested simulation levels with progressively finer spatial resolution. The hierarchy is designed to connect statistical samples of dwarf galaxies across a broad range of environments to increasingly detailed simulations of their internal properties. \darwini\ is the first level: a full cosmological box with a box length of $L_{\rm box}=65.536$~cMpc and a best spatial resolution of 500~pc. Although its resolution is relatively coarse with respect to the typical size of dwarf galaxies, the large volume provides a statistical sample of dwarf galaxies spanning diverse cosmological environments and defines the parent population from which subsequent zoom-in targets are selected. \darwinii\ comprises zoom-in simulations of 24 fields and a Milky way analog with its associated major filament, with a minimum cell size of 62.5~pc, targeting diverse environments, including field, filament, and Milky Way-like. These \darwinii\ simulations enable us to investigate how the global properties of dwarf galaxies depend on their large-scale environments. 
The \darwini\ volume serves as a parent sample from which additional \darwinii\ targets can be selected in future studies for a broad range of applications.

Because high resolution is restricted to selected regions and galaxies, \darwin\ cannot simultaneously provide a large  statistical sample and high numerical resolution for the same population. To mitigate this limitation, we use machine-learning techniques to infer high-resolution properties for the low-resolution galaxies in \darwini. The model is trained on galaxies matched between \darwini\ and \darwinii, following the approach commonly referred to as ``galaxy painting'' \citep{Kamdar2016, 2019MNRAS.489.3565J,Villaescusa-Navarro21, 2024ApJ...965..156J}. 

\begin{figure*}
\centering 
\includegraphics[width=\linewidth]{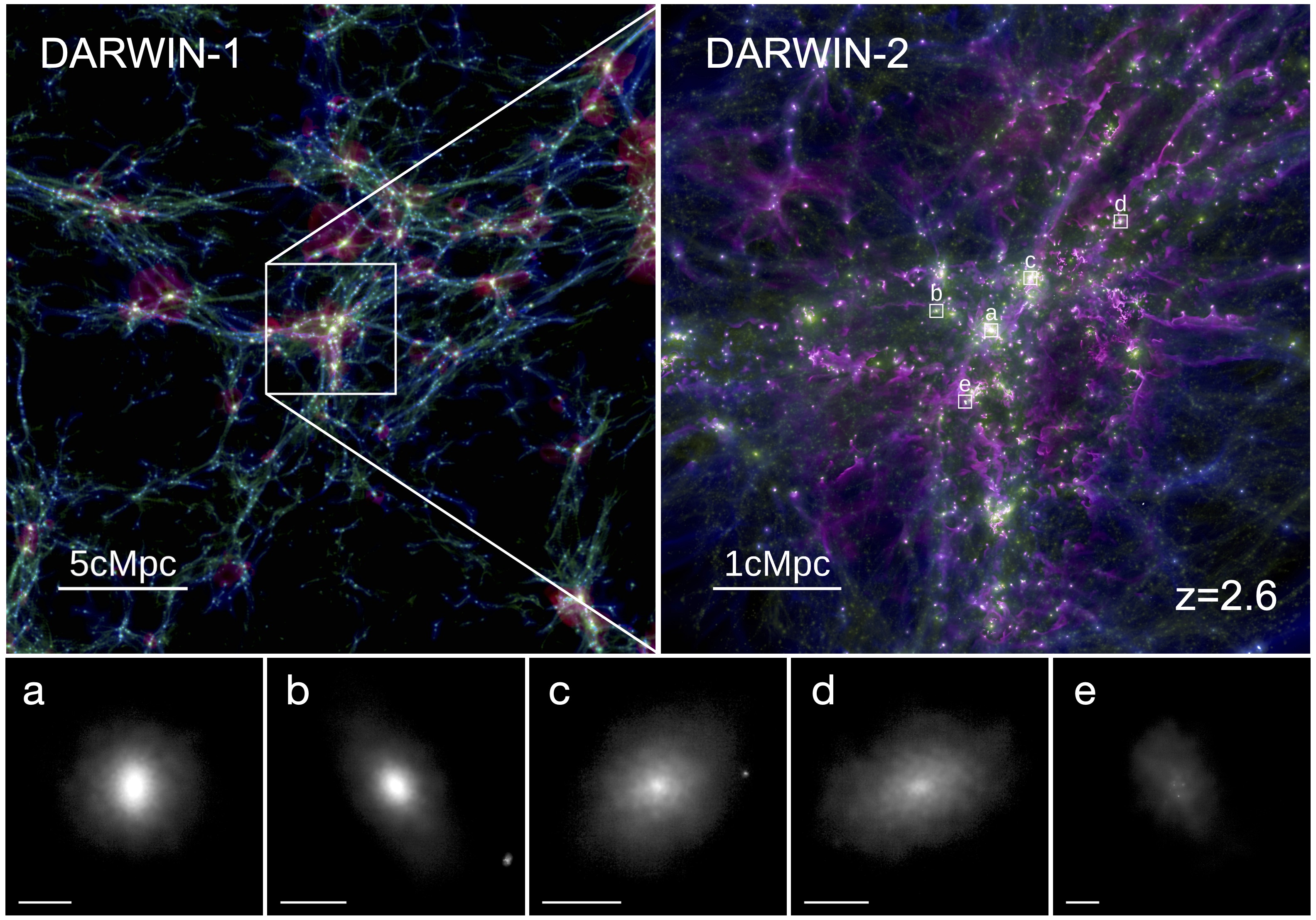}
\caption{Color-composite images of a region in \darwini\ centered on a
Milky Way--like group (left) and its high-resolution counterpart in
\darwinii\ (right) at $z=2.6$. Red, green, and blue represent thermal
energy, dark matter density, and gas density, respectively. Bottom panels show the stellar mass density maps of the five galaxies indicated by white squares in the top right panel. The white scale bar marks 2~ckpc in each bottom panel.} 
\label{fig:darwin12}
\end{figure*}

This paper presents the design, numerical implementation, and general performance of the \darwin\ simulations, together with the data products available. We focus in this paper on establishing the \darwin~ framework and assessing how well it reproduces observed global galaxy properties, rather than on individual science questions. Detailed studies of specific science topics will be presented in forthcoming papers. The remainder of the paper is organized as follows. In \S~\ref{sec:simulations}, we describe the simulation setup, including the numerical code, cosmological parameters, initial conditions, baryonic physics, optimization procedures, galaxy identification, and merger tree construction. \S~\ref{sec:results} presents the global galaxy properties and scaling relations of the \darwin\ simulations. \S~\ref{sec:products} describes the data products, including mock observations and machine-learning-based galaxy painting. Finally, we summarize our main results and discuss their implications in \S~\ref{sec:summary}. 

\section{Simulations}\label{sec:simulations}

The \darwin\ simulations comprise three distinct sets with complementary scientific goals. \darwini\ consists of a large-volume cosmological run that samples galaxy populations across diverse environments at relatively low resolution. The latest available snapshot of the hydrodynamical \darwini\ simulation
is at $z=0.19$. In contrast, \darwinii\ comprises a suite of multi-zoom-in simulations targeting field environments and a Milky Way analog at tens-of-parsec resolution with more detailed  baryonic physics. Figure~\ref{fig:darwin12} illustrates a region in \darwini\ containing a Milky Way analog (top left), its high-resolution counterpart in \darwinii\ (top right), and five massive galaxies with diverse morphologies drawn from the same region (bottom). Table~\ref{tab:sim_comparison} summarizes the main differences in numerical specifications between the two simulation suites. \darwiniii\ is designed to resolve the multiphase interstellar medium (ISM) and circumgalactic medium (CGM) of individual dwarf galaxies at substantially higher resolution and will be presented in a separate paper. In this section, we describe the numerical methods and initial conditions adopted for \darwini\ and \darwinii.

\subsection{Code}\label{sub:code} 

To simulate the formation and evolution of dwarf galaxies, we use the adaptive mesh refinement (AMR) Eulerian hydrodynamics code \ramses~\citep{teyssier02}. Our implementation builds on the public version available through the Git repository\footnote{https://bitbucket.org/rteyssie/ramses}, with modifications for cosmological galaxy formation simulations, including the treatment of active galactic nucleus physics, star formation, stellar feedback, and chemical evolution, as described below. To enable large-volume simulations on NURION, where available memory is limited to 1.3\,GB per CPU core\footnote{https://docs-ksc.gitbook.io/nurion-user-guide-eng/system/specifications-and-configuration}, we have developed a hybrid MPI+OpenMP version that reduces the memory footprint while maintaining parallel scalability. The hybrid parallelization strategy is described in Appendix~\ref{sub:omp}.

\ramses\ solves the Euler equations for gas dynamics using a second-order Godunov scheme based on the MUSCL–Hancock method, with a minmod slope limiter and an approximate Harten–Lax–van Leer–Contact (HLLC) Riemann solver~\citep{toro94}. The gravitational potential is computed using a particle–mesh method with cloud-in-cell interpolation, solving the Poisson equation with a multigrid method on coarse AMR levels~\citep{guillet11} and a conjugate gradient method on finer levels~\citep{teyssier02}. The AMR hierarchy is organized as a fully threaded oct-tree structure \citep{khokhlov98}, with cells refined when their mass exceeds a prescribed threshold, following a quasi-Lagrangian criterion. This scheme concentrates computational resolution in dense, self-gravitating gas while limiting excessive refinement in low-density regions. In the \darwinii\ simulations, we additionally include radiative transfer using the two-moment formulation with a Global Lax–Friedrich flux function and an M1 closure for the Eddington tensor with a reduced speed of light ratio 0.01~\citep{Rosdahl13}, enabling a coupled evolution of radiation and gas dynamics within the AMR hierarchy. Details of the individual simulation setups are given below.


\begin{deluxetable}{ccc}
\tablewidth{0.48\textwidth}
\tablecaption{\darwini\ and \darwinii\ comparison}
\label{tab:sim_comparison}

\tablehead{
\colhead{} & \colhead{\darwini} & \colhead{\darwinii}
}

\startdata
$\Delta x$ (proper) & 500~pc & 62.5~pc \\
$m_{\rm DM}$ & $8.9\times10^{6}\,\msun$ & $1.7\times10^{4}\,\msun$ \\
$m_\star$ & $10^{5}\,\msun$ & $10^{4}\,\msun$ \\
$m_{\rm BH,seed}$ & $5\times10^{4}\,\msun$ & $2.5\times10^{4}\,\msun$ \\
Chemical elements & H, O, Fe & H, N\tablenotemark{a}, O, Mg, Fe \\
Ion Species & - & \ion{H}{2}, \ion{He}{2}, \ion{He}{3} \\
\enddata
\tablenotetext{a}{\darwinii\ additionally tracks magnesium and, in some runs, nitrogen. See Table~\ref{tab:sim_list} for more details.}
\end{deluxetable}

\subsection{Cosmology and Initial Conditions}\label{sub:ic} 

We adopt a $\Lambda$CDM cosmology with the parameters compatible with the {\it Planck} measurements \citep{planck18}: matter density $\Omega_{\rm m}$=0.315, dark energy density $\Omega_{\Lambda}$=0.685, baryon density $\Omega_{\rm b}$=0.045, amplitude of matter density fluctuations $\sigma_8=0.812$, scalar spectral index $n_s=0.965$, and Hubble constant $H_0=67.4\,{\rm km\,s^{-1}\,Mpc^{-1}}$. The linear matter power spectrum is calculated using the \texttt{CAMB} package~\citep{lew00}. The initial conditions at $z=200$ are generated with \texttt{MUSIC} \citep{hah11} using second-order Lagrangian perturbation theory \citep{sco98,lhu14}. 

For \darwini, the initial density field is generated in a periodic box with a side length of 65.536~cMpc using a $1024^3$ base grid (level=10) and the same number of dark matter (DM) particles with $m_{\rm DM}=8.9\times10^6\,\msun$. Cells are adaptively refined using a quasi-Lagrangian criterion: a cell is refined when its mass exceeds eight times the DM mass resolution, down to a minimum physical cell size of 0.5~kpc (level=17). To maintain an approximately constant minimum cell size ($\Delta x=0.5~$kpc) within the comoving box, a new refinement level is introduced at expansion factors $a_{\rm exp}\approx0.0125$, $0.025$, $0.05$, $0.1$, $0.2$, $0.4$, and $0.8$. The time step is adaptive, with a shared step size at each refinement level, and varies by a factor of 2 between adjacent levels. The time step is set by the Courant condition, with a Courant factor of 0.8.

\darwinii\ comprises a suite of zoom-in simulations targeting regions selected from the \darwini\ DM-only simulation. The surrounding low-resolution volume is initialized on a level-8 base grid, rather than the level-10 grid used in the original \darwini\ run, to reduce the computational cost. Within the zoom-in regions, the effective initial resolution corresponds to level 13. We follow the same refinement strategy as in \darwini, increasing the maximum refinement level at the same expansion factors up to level 20. This maintains an approximately constant proper spatial resolution of $\Delta x=62.5$~pc, eight times finer than that of \darwini. Refinement is restricted to the moving zoom-in regions using two conditions. First, we introduce a passive scalar initialized to unity within the zoom-in regions and zero elsewhere, and allow refinement only where its value exceeds 0.9. Second, only high-resolution DM particles contribute to the quasi-Lagrangian refinement criterion.

\begin{figure*}
\centering 
\includegraphics[width=\linewidth]{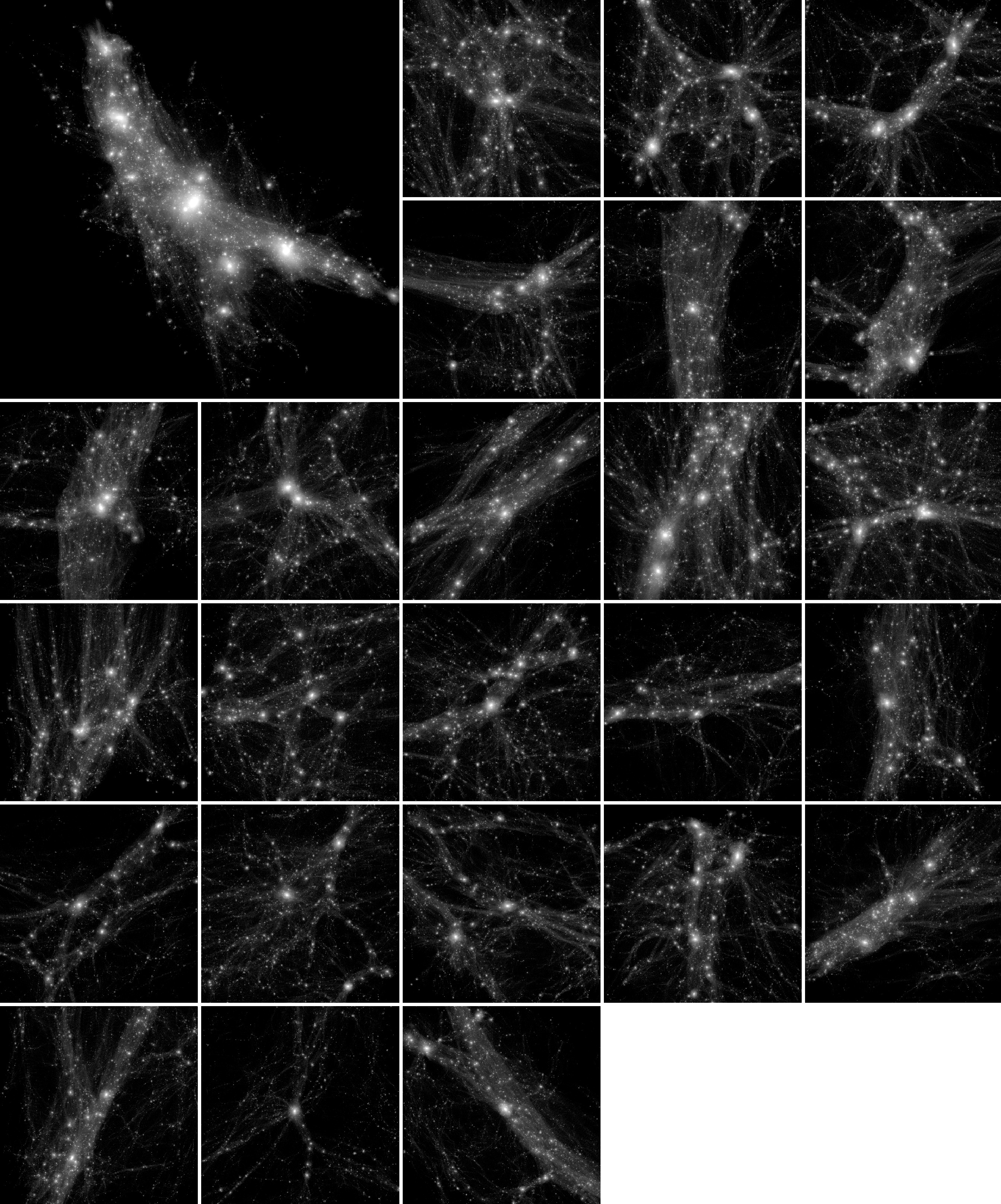}
\caption{Dark matter distribution in the DM-only \darwinii\ run at $z=0$. The top-left panel shows a Milky Way analog connected to a filament, whereas the other panels show 24 field regions. Each small panel has a physical side length of 4.5~Mpc; the top-left panel is twice as large. Only the highest-resolution particles are shown to illustrate the target structures in the zoomed regions.} 
\label{fig:darwin2_dm_map}
\end{figure*}

To investigate the evolution of dwarf galaxies in different environments, the \darwinii\ suite includes high-resolution regions centered on galaxy groups, massive galaxies, and isolated field galaxies. For regions centered on groups or massive galaxies, we define a spherical high-resolution volume extending to  $3.5\,R_\mathrm{200c}$ at $z=0$, where $R_\mathrm{200c}$ is the radius within which the mean enclosed density is 200 times the critical density of the Universe. Where necessary, we extend this region along the major filaments connected to the central system out to $10\,R_\mathrm{200c}$, using a cylindrical radius of $3.5\,R_\mathrm{200c}$. In contrast, the field regions are selected to contain only halos with $\mhalo < 5\times10^{11}~\msun$. Each field region is defined as a (4~cMpc)$^3$ cubic volume in which all dwarf galaxies lie at least 7~cMpc from massive galaxies or groups and 10~cMpc from galaxy clusters. These criteria minimize the influence of nearby massive structures and provide a sample of dwarf galaxies in relatively isolated environments. For computational efficiency, each \darwinii\ field run combines up to five zoom-in regions within a single simulation volume. The current \darwinii\ suite includes one Milky Way-like galaxy and its associated major filaments, along with 24 field regions. Figure~\ref{fig:darwin2_dm_map} shows the dark matter distribution at $z=0$ in the DM-only \darwinii\ run, highlighting the Milky Way analog and the 24 field regions. The properties of all \darwinii\ zoom-in regions are summarized in Table~\ref{tab:sim_list} in Appendix~\ref{sec:sim_list}.

\subsection{Subgrid Physics}\label{sub:subgrid} 
\subsubsection{Gas cooling and heating}\label{sub:cooling} 

\darwini\ and \darwinii\ include radiative cooling and heating from both primordial and metal species. Primordial cooling processes include collisional ionization, collisional excitation, recombination, Bremsstrahlung, and Compton cooling/heating, following the standard implementation in \ramses\. Metal cooling above a temperature $T\sim10^4$\,K is computed using pre-tabulated {\sc cloudy} cooling and heating rates based on the `cc07' model distributed with \ramses~\citep{ferland98}. At lower temperatures, metal-enriched gas cools further through fine-structure line emission using the fitting formula of \citet{koyama02}.

The treatment of primordial chemistry differs between the two simulations. In \darwini, cooling and heating rates are computed using pre-tabulated {\sc cloudy} tables under the assumption of ionization equilibrium with a spatially uniform UV background~\citep{haardt96}, following the default \ramses\ implementation. In contrast, \darwinii\ self-consistently follows the non-equilibrium ionization states of hydrogen and helium, coupled to three radiation bins defined by the ionization thresholds of \ion{H}{1}, \ion{He}{1}, and \ion{He}{2}. For \darwinii, we also adopt the updated UV background model of \citet{haardt12} and attenuate it using a self-shielding prescription,  $\exp(-n_{\rm H}/0.01\,\mathrm{cm^{-3}})$, following \citet{rosdahl12}. In both simulations, the spatially uniform UV background is activated at $z_{\rm reion}=8.5$.

\subsubsection{Star formation}\label{sub:SF} 

In our simulations, the star formation rate follows a Schmidt law~\citep{schmidt59}:
\begin{equation}
\label{eqn:schmidt}
\frac{{\rm d}\rho_{\star}}{{\rm d}t}=\varepsilon_{\rm ff}\frac{\rho_{\rm gas}}{t_{\rm ff}},
\end{equation}
where $\rho_{\rm gas}$ is the gas density, $t_{\rm ff}=\sqrt{(3\pi/32G\rho_\mathrm{gas})}$ is the free-fall time, $G$ is the gravitational constant, and $\varepsilon_\mathrm{ff}$ is the local star formation efficiency per free-fall time.

In \darwini, we adopt a fixed star formation efficiency of $\varepsilon_{\rm ff}=1$. Star formation is permitted in grid cells with a hydrogen number density $n_{\rm H}>0.1\,{\rm cm^{-3}}$ and a converging velocity field ($\vec{\nabla} \cdot \vec{v}<0$), following \citet{cen92}. The minimum stellar particle mass is set to $M_{\rm star,min}=10^5\,\msun$. Stellar particles are formed with masses $m_{\rm star} = N_{\rm star}\times M_{\rm star, min}\,$, where $N_{\rm star}$ is an integer multiple drawn stochastically from a Poisson distribution such that the mean star formation rate follows the Schmidt law~\citep{rasera06}. To prevent excessive gas depletion from individual cells,  the mass of a newly formed star particle is capped at 90\% of the gas mass in the host cell.

\darwinii\ adopts a minimum grid size of $\sim60$~pc, which marginally resolves a multiphase ISM. Motivated by studies showing that $\varepsilon_\mathrm{ff}$ is not constant but varies with gas properties \citep{padoan11,federrath12}, we compute $\varepsilon_\mathrm{ff}$ using a gravo-turbulent model \citep{kimm17,trebitsch17}, as
\begin{equation}
\label{eqn:eff}
\varepsilon_{\rm ff}=\frac{\epsilon_{\rm ecc}}{2\phi_{\rm t}} \exp \bigg( \frac{3}{8} \sigma^2_s \bigg) \bigg[ 1+ {\rm erf} \bigg( \frac{\sigma^2_s-s_{\rm crit}}{\sqrt{2\sigma^2_s}}\bigg) \bigg].
\end{equation}
Here, $\epsilon_\mathrm{ecc}\approx0.5$ is the fraction of gas that can be accreted onto stars after accounting for mass loss from protostellar jets and outflows, and $1/\phi_{\rm t}\approx0.49$ is a numerical factor that accounts for the multi-free-fall nature of star-forming clouds~\citep{hennebelle11}. The variance of the logarithmic density contrast, defined as $s\equiv\ln(\rho/\rho_0)$, is given by $\sigma^2_s= \ln (1+b^2\mathcal{M}^2)$, where $b$ depends on the turbulence driving mode. We adopt $b=0.4$, corresponding to a mixture of solenoidal and compressive driving. The quantity $\mathcal{M}$ is the sonic Mach number, and $\rho_0$ is the mean density of the cloud. The critical overdensity above which gas becomes gravitationally unstable, $s_{\rm crit}$, is approximated following \citet{federrath12} as
\begin{equation}
\label{eqn:scrit}
s_{\rm crit}=\ln(0.067\theta^{-2}\alpha_{\rm vir}\mathcal{M}^2),
\end{equation}
where $\theta=0.97$ is a numerical factor accounting for the uncertainty in the post-shock thickness relative to the cloud size, and $\alpha_{\rm vir}=2E_{\rm kin}/|E_{\rm grav}|\approx 5(\sigma_{\rm gas}^2+c_s^2)/\pi \rho_{\rm gas} G \Delta x^2$ is the virial parameter of the cloud, where $c_s$ is the gas sound speed. We estimate the local Mach number from the velocity dispersion between neighboring grid cells, defined as $\sigma_{\rm gas}^2={\rm Tr} \left( \vec{\nabla} v^{\rm T} \vec{\nabla} v \right) \Delta x^2$, where $v$ is the divergence- and rotation-free local velocity field. This prescription favors star formation in regions where thermal and turbulent support is insufficient to prevent gravitational collapse. Since \darwinii\ has a spatial resolution three levels finer than that of \darwini, we allow star formation in grid cells with $\nH>6.4\,\cmq$, based on the scaling of the Jeans density as $\Delta x^{-2}$. Once $\varepsilon_{\rm ff}$ is determined, the stellar particle mass is sampled stochastically from a Poisson distribution with a minimum mass of $M_{\rm star,min}=10^4\,\msun$, again following \citet{rasera06}.  

\subsubsection{Stellar feedback and chemical enrichment}\label{sub:feedback} 

We include stellar feedback from Type II and Type Ia supernovae (SNe) in both \darwini\ and \darwinii. Core-collapse SNe from massive ($M\geq 8\msun$) stars are modeled as a single event per star particle using the mechanical SN feedback scheme of \citet{kimm14}. We stochastically assign the explosion time of Type II SNe using a fitting function constructed from the Kroupa IMF with slopes of $-1.3$ for stellar masses of $0.1$-$0.5\,\msun$, and $-2.35$ for masses of $0.5$-$100\,\msun$~\citep{kroupa01} and the corresponding mass-dependent stellar lifetimes, which maps a uniform random deviate onto an explosion time. The model accounts for whether the SN remnant is in the energy-conserving or momentum-conserving phase and injects the appropriate radial momentum, approaching the terminal value,
\begin{equation}
    p_{\rm SNII}=p_{\rm rad}\, n_{\rm H}^{-2/17} (N_{\rm SNII}\,E_{51})^{16/17}\,Z'^\alpha,
\end{equation}
where $p_{\rm rad}=2.5\times10^5\,\msun\,\kms$ is the terminal momentum per explosion at $\nH=1\,\cmq$ and solar metallicity \citep[e.g.,][]{thornton98}, $n_{\rm H}$ is the hydrogen number density in units of $\cmq$, $N_{\rm SNII}$ is the number of Type II SNe in a cell, $E_{51}$ is the explosion energy in units of $10^{51}\,$erg, and $Z'\equiv \max\left(Z_{\rm gas}/Z_{\odot},0.01\right)$ is the gas metallicity normalized by the solar value. To mitigate numerical overcooling \citep[e.g.,][]{rosdahl17}, the canonical SN explosion energy of $10^{51}\,$erg is boosted by a factor of 7 at solar metallicity in \darwini\ and \darwinii, while  metallicity exponents of $\alpha=0.15$ in \darwini\ and 0.42 in \darwinii\ are adopted as part of the calibration of stellar mass growth at $z=0$. The feedback boost is weakened toward lower metallicity because such conditions often occur at high redshift, where star formation is more active, or in systems with shallower gravitational potential wells; applying the same enhancement in these regimes would suppress stellar mass growth excessively. Accordingly, at $Z_{\rm gas}/Z_{\odot}=0.01$, the boosted energy is reduced to one-half of the initial boosted value in \darwini\ and one-seventh in \darwinii. 

Type Ia SNe are implemented as thermal feedback, by a stochastically sampling a delay-time distribution following the power-law form proposed by \citet{maoz12}:
\begin{equation}
    \Psi(t)=\phi \,t^{-1},
\end{equation}
where $\phi$ is the normalization constant, determined by the time-integrated number of SNe Ia per unit stellar mass of a simple stellar population (SSP),
\begin{equation}
    N_{\rm SNIa}/M_{\rm SSP}=\int_{t_0}^{t_1} \Psi(t) dt,
\end{equation}
with $t_0=0.05$~Gyr and $t_1=13.7$~Gyr. We adopt   $\phi=2.35\times10^{-4}\,\msun^{-1}$, corresponding to $N_{\rm SNIa}/M_{\rm SSP}\approx1.3\times10^{-3}\,\msun^{-1}$. 

With radiative transfer implemented in \darwinii, young massive stars produce Lyman continuum photons that ionize hydrogen and heat the gas to $T\sim 10^4\,{\rm K}$. This process creates overpressurized H\textsc{ii} regions that expand into the surrounding medium \citep[e.g.,][]{krumholz07,walch12,dale14}. Absorbed ionizing photons also transfer momentum to the neutral ISM or dust \citep[e.g.][]{haehnelt95}, providing additional channels of stellar feedback. We use three photon groups with energies of $E_1=13.6-24.59\,$eV, $E_2=24.59-54.42\,$eV, and $E_3>54.42\,$eV, corresponding to the ionization of neutral hydrogen, neutral helium, and singly ionized helium, respectively. The photon production rates of stellar populations are computed using the Binary Population and Spectral Synthesis model~\citep[][v2.2.1]{eldridge17,stanway18} assuming the Kroupa IMF~\citep{kroupa01}. 

We use separate chemical yield tables for Type Ia SNe~\citep{seitenzhal13} and Type II SNe. For Type II SNe, we adopt the yield tables implemented in \textsc{Starburst99}~\citep{leitherer99}, which are based on the nucleosynthetic yields of \citet{woosley95}. The ejecta masses are determined from the adopted IMF. In \darwini, we trace the abundances of hydrogen, oxygen, iron, and total metals, whereas \darwinii\ additionally tracks magnesium and nitrogen. We initialize the metallicity to $10^{-3}\,Z_\odot$, with the relative elemental abundances following the solar abundance pattern of \citet{asplund09}.


\subsubsection{Formation and evolution of massive black holes}\label{sub:mbh} 

We employ the black hole (BH) seeding, growth, and AGN feedback schemes formulated by \citet{dubois21}. BHs are seeded with an initial mass of $5\times10^4\,\msun$ (\darwini) or $2.5\times10^4\,\msun$ (\darwinii) in cells with hydrogen number densities above $\nH=0.1\,\cmq$ (\darwini) or $\nH=1.6\,\cmq$ (\darwinii). A BH is seeded only if no other BH is present within 10~ckpc in \darwini\ or 5~ckpc in \darwinii.

At the resolution of cosmological simulations, seed-mass BHs may not efficiently sink toward galaxy centers through dynamical friction~\citep[e.g.,][]{ma21,bahe22}. Subgrid models are therefore commonly used to account for unresolved drag from gas, stars, or DM~\citep{Dubois13drag,tremmel15,biernacki17,pfister19,ogiya20,lescaudron23,mukherjee25,zhou25,li25,shin25}. In \darwini, we include gas dynamical friction following~\citet{Dubois13drag}:
\begin{equation}
\boldsymbol{F}_{\rm gDF}=-4\pi \alpha_{\rm gas} {f}_{\rm gas} \rho \frac{G^2M_{\rm BH}^2}{c_{\rm s}^2}\frac{\boldsymbol{u_{\rm BH}}}{u_{\rm BH}},
\end{equation}
where $\rho$ is the gas density, $f_{\rm gas}$ is a Mach-number-dependent factor~\citep{Ostriker99}, with $\mathcal{M}=u_{\rm BH}/c_{\rm s}$, $u_{\rm BH}$ and $c_{\rm s}$ are the BH-gas relative velocity and gas sound speed, respectively, and $M_{\rm BH}$ is the BH mass. In regions above the BH seeding threshold, the drag force is boosted by a factor $\alpha_{\rm gas}=\max[(\rho/\rho_{\rm crit})^\beta,1]$, where $\rho_{\rm crit}$ is the corresponding gas density threshold. We adopt $\beta=2$ for \darwini\ and $\beta=1$ for \darwinii. 

In \darwinii, we additionally include dynamical friction from stars and DM particles, following \citet{tremmel15} as modified by \citet{pfister19}:
\begin{multline}
\boldsymbol{F}_{\rm pDF}=-4\pi \alpha_{p} G^2M_{\rm BH}^2\frac{\boldsymbol{\bar{v}}_{\rm BH}}{\bar{v}_{\rm BH}^3} \bigg\{ \rho_{s}\ln  \Lambda \\
+\frac{1}{V} \sum\limits_{i\in \mathcal{P}_f}  m_i \left[\ln \left(\frac{v_i+v_{\rm BH}}{v_i-v_{\rm BH}} \right) 
-2\frac{v_{\rm BH}}{v_i} \right] \bigg\},
\end{multline}
where $\boldsymbol{\bar{v}}_{\rm BH}$ is the BH velocity relative to the mass-weighted velocity of background particles within a spherical region of volume $V$ and radius $r<2\Delta x$ centered on the BH, $\rho_s$ denotes the density of background particles moving more slowly than the BH, and $\ln \Lambda\sim \ln (b_{\rm max}/b_{\rm min})$ is the Coulomb logarithm. We adopt $b_{\rm max}=2\Delta x$ and $b_{\rm min}={\rm max}(GM_{\rm BH}/v^2_{\rm BH},R_{\rm Sch})$, where $R_{\rm Sch}$ is the Schwarzschild radius. The contribution from particles with velocities faster than the BH is computed explicitly using their individual masses $m_i$ and relative velocities $v_i$ for all particles in $\mathcal{P}_f$. The dynamical friction is further boosted by $\alpha_p=\max[\bar\rho_p/\rho_{\rm crit},1]$, where $\bar\rho_p$ is the mean particle density within the spherical region. 
The contributions from stellar and DM particles are computed separately.

BHs grow by accreting surrounding gas at the Bondi-Hoyle-Lyttleton rate, $\dot M_{\rm BH}=(1-\varepsilon_{\rm r})\dot M_{\rm BHL}$, where $\varepsilon_{\rm r}$ is the spin-dependent radiative efficiency and 
\begin{equation}
\dot M_{\rm BHL}=4\pi\bar \rho \frac{ G^2M_{\rm BH}^2}{(\bar u^2+ \bar c_{\rm s}^2)^{3/2}}.
\end{equation}
The accretion rate is capped at the Eddington limit, 
\begin{equation}
\dot M_{\rm Edd}=4\pi G \frac{M_{\rm BH}m_{\rm p}}{\varepsilon_{\rm r}\sigma_{\rm T} c},
\end{equation}
where $c$ is the speed of light and $\sigma_{\rm T}$ is the Thomson cross-section. Quantities marked with bars are kernel-weighted averages of local gas properties~\citep[see][for more details]{Dubois12}. 

BH pairs merge when their separation falls below $4\Delta x$, where $\Delta x$ is the minimum cell size at a given epoch, and their relative velocity is below the binary escape speed.

\subsubsection{AGN feedback}\label{sub:agn} 

AGN feedback is implemented using dual jet-heating modes that depend on the Eddington ratio $\chi=\dot M_{\rm BH}/\dot M_{\rm Edd}$~\citep{Dubois12}. At low ($\chi\le0.01$) and high ($\chi>0.01$) accretion rates, feedback operates in the radio (jet) and quasar (thermal) modes, respectively. 

In the thermal mode, we adopt a coupling efficiency of $\epsilon_{\rm f,h}=0.6$ to reproduce the $M_{\rm BH}$--$M_\star$ relation in both \darwini\ and -2.
The corresponding energy injection rate is given by $\dot E_{\rm BH,h}=\varepsilon_{\rm r}\varepsilon_{\rm f,h}\dot M_{\rm BHL} c^2$.
BH spins are evolved self-consistently through binary BH coalescence \citep[following][]{Rezzolla08} and smooth gas accretion \citep[see][for details]{Dubois14spin}. A geometrically thin, radiatively efficient accretion disk~\citep{Shakura73} is assumed in the quasar mode.

In the jet mode, we adopt the magnetically choked accretion flow solution of \citet{mckinney12} for BH spin evolution. The energy injection rate in the jet mode is given by $\dot E_{\rm BH,j}=\varepsilon_{\rm f,j}\dot M_{\rm BHL} c^2$ with the energy coupling efficiency of the jet mode, where the jet coupling efficiency $\varepsilon_{\rm f,j}$ is taken from \cite{mckinney12}. This efficiency has a U-shaped dependence on BH spin, reaching a minimum of a few per cent for non-rotating BHs and around 100\% for maximally spinning ones (see \citealp{dubois21} for details). Jets deposit momentum and energy into a bipolar outflow aligned with the BH spin with zero opening angle and redistribute mass with a constant mass loading factor of $\eta=\dot M_{\rm J}/\dot M_{\rm BH}=100$.


\subsection{Structure identification and merger trees}\label{sub:PgalF} 

We use a galaxy finder \pgalf~\citep{kim23} to identify self-bound objects in snapshots from the \darwin\ simulations. First, \pgalf\ identifies Friends-of-Friends (FoF) groups from the distribution of heterogeneous matter components in \darwin, including stars, MBHs, dark matter particles, and gas cells. For this mixture of particle types and masses, \pgalf\ assigns an adaptive linking length to each particle. A pair of particles with individual linking lengths $l_1$ and $l_2$ is linked when their separation is smaller than the arithmetic mean of the two linking lengths, $(l_1+l_2)/2$. It then identifies self-bound substructures within the FoF
halos.

We classify a substructure as a galaxy if it contains stellar particles. To find galaxies within an FoF halo, \pgalf\ first constructs a three dimensional uniform-grid stellar density field and hierarchically determines the membership of particles bound to galaxies centered on stellar density peaks. A bound particle is assigned to a galaxy if it lies within the tidal radius of the galaxy. We note that a galaxy identified in this process generally consists of heterogeneous matter components. For substructures without stellar particles, a similar procedure is applied to the remaining matter species. For a full description of the method, we refer to \cite{kim23}.

We construct galaxy and subhalo merger trees by tracing progenitor--descendant branches between adjacent snapshots, using stellar particles for galaxies and dark matter particles for subhalos. For galaxies, we use a core-based branch-tracing scheme which identifies a branch using the central stellar component. This choice reduces spurious associations caused by stellar particles in galaxy outskirts, which are more susceptible to close-pair contamination, tidal features, and limitations of the galaxy finder. We enforce a non-fragmentation condition, so that two galaxies at the same snapshot cannot share the same main progenitor. When adjacent-snapshot matching fails this consistency condition because of a fly-by event or a temporary failure of structure identification, we revisit earlier snapshots and reconnect the branch to an appropriate progenitor candidate. Details of the tree building scheme and correction procedure are provided in Appendix~\ref{sec:appendix-tree}.

\section{Global Properties and Scaling Laws}\label{sec:results} 
In this section, we present the global galaxy properties and scaling relations of the \darwin\ simulations. We first examine the evolution of the cosmic star formation rate density in \darwini\ (Section~\ref{sub:sfrd}) and the galaxy stellar mass--halo mass relation (Section~\ref{sub:stellar-halo}). We then present the evolution of the galaxy stellar mass function in \darwini\ (Section~\ref{sub:gsmf}), the galaxy mass--size relation (Section~\ref{sub:mass-size}), and the gas-phase metallicity at high redshift and stellar metallicity in the local universe (Section~\ref{sub:metallicity}). We also investigate the relation between cold-gas and star formation rate surface densities in \darwinii\ (Section~\ref{sub:gas-sfr}) and the black hole mass--galaxy stellar mass relation in both simulations (Section~\ref{sub:bh-stellar}). Together, these diagnostics allow us to assess how well the \darwin\ simulations reproduce the main observed properties and scaling relations of galaxies across cosmic time.

\subsection{Star formation rate density evolution of \darwini}\label{sub:sfrd}

\begin{figure}
\centering 
\includegraphics[width=\linewidth]{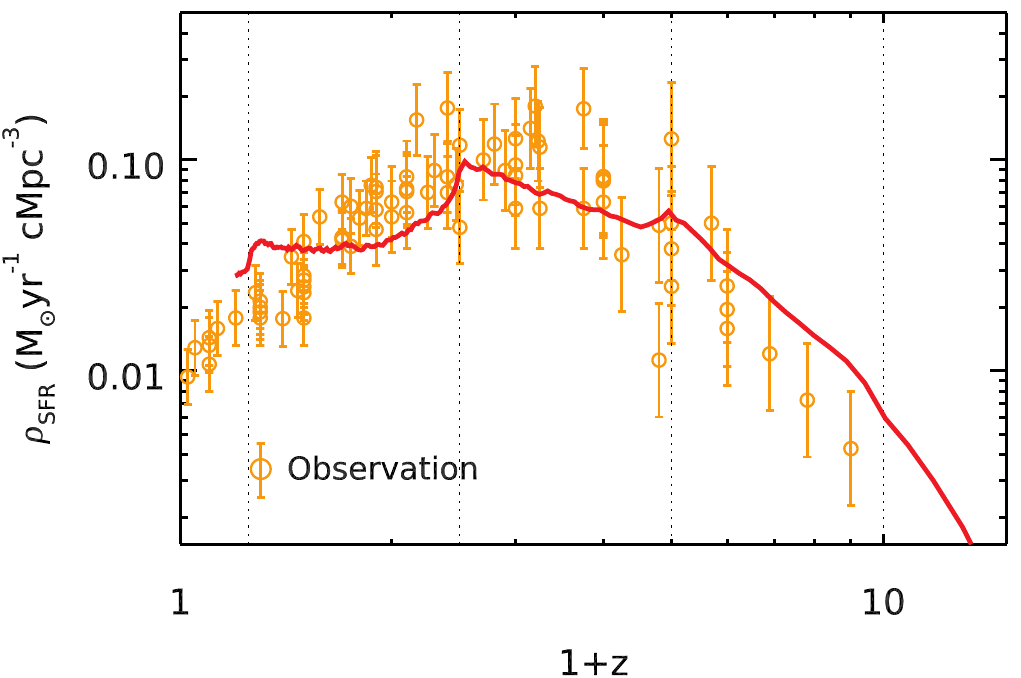}
\caption{Star formation rate density evolution in \darwini. Orange symbols with error bars illustrate the observational data compiled by \citet{behroozi13}. \darwini\ follows the observed trend well at $z>1$, but the simulated SFRD declines more slowly toward lower redshift. The vertical dotted lines mark the epochs when the maximum grid level increases to maintain approximately constant physical resolution.}
\label{fig:sfrd}
\end{figure}

The star formation rate density (SFRD) is shown in Figure~\ref{fig:sfrd} as a function of redshift. We reconstruct the cosmic star formation history from the initial masses and birth times of all stellar particles in the simulation volume at the $z=0.25$ snapshot. We bin cosmic time uniformly in proper units ($\Delta t=50$~Myr), measure the stellar mass formed in each time interval.

We compare the \darwini\ SFRD with the compilation presented by \citet{behroozi13}. The model SFRD follows the observed trend reasonably well up to the peak epoch of cosmic star formation, at $z\simeq1.5-2$, but declines more slowly thereafter. This produces an excess at low redshift, indicating that star formation remains too active in the simulation. Several effects could contribute, including excessive gas supply or recycling, overcooling in massive halos, and prolonged star formation in low-mass systems. As discussed below, the elevated low-$z$ SFRD is  qualitatively consistent with the overabundance of massive galaxies in the low-redshift galaxy stellar mass function. Determining which galaxy or halo mass range drives the discrepancy, however, requires decomposing the SFRD by host mass.

In this calculation, we include all stellar particles, not only those bound to galaxies. At $z=0.25$, more than 91\% are bound to galaxies, about 8\% are unbound but remain within FoF halos, and less than 0.5\% reside in a diffuse component outside any FoF halos. EEven at $z>5$, the fraction of stellar particles outside FoF halos remains below 2\%. Excluding this component would lower the normalization by less than $0.01$~dex, making its contribution negligible for the SFRD measured here.

We note that \citet{behroozi13} adopt a Chabrier IMF, whereas \darwini\ uses a Kroupa IMF, which typically yields an SFRD higher by $\sim0.05$~dex than that based on a Chabrier IMF at fixed observed luminosity. Converting the \darwini\ values to a Chabrier IMF would therefore modestly reduce the apparent low-$z$ excess. The offset, however, is much smaller than the observational scatter and does not affect the redshift-dependent trend discussed here.

In \ramses, the maximum refinement level is increased at $z=19$, 9, 4, 1.5, and 0.25 to maintain an approximately constant physical resolution. These discrete refinement transitions can introduce artificial features in the star formation history particularly noticeable at $z=1.5$, near cosmic noon. Since we adopt a local converging-flow criterion in the star formation prescription of \darwini, these transitions do not produce an abrupt increase of star formation rates as seen in Figure~\ref{fig:sfrd}. However, they can still cause transient numerical artifacts in the gas properties through the sudden appearance of better-resolved dense structure (see G. Lee et al. in prep). 
 
Several efforts have been made to mitigate these artifacts, including a smoothing transition scheme based on a Jeans-length criterion~\citep{han26} and a scheme that allows additional refinement of dense regions at any epoch while imposing a Jeans pressure floor, albeit at increased computational cost because more cells are refined (G. Lee et al. in prep). 
G. Lee et al. (in prep) find that the total amount of stars formed over cosmic time does not significantly differ between the control simulation and simulations employing the new refinement scheme. However, abrupt global refinement without a smooth transition appears to have a more noticeable effect on galaxy morphology.

\subsection{Galaxy stellar mass -- halo mass relation}\label{sub:stellar-halo}

\begin{figure}
\centering 
\includegraphics[width=\linewidth]{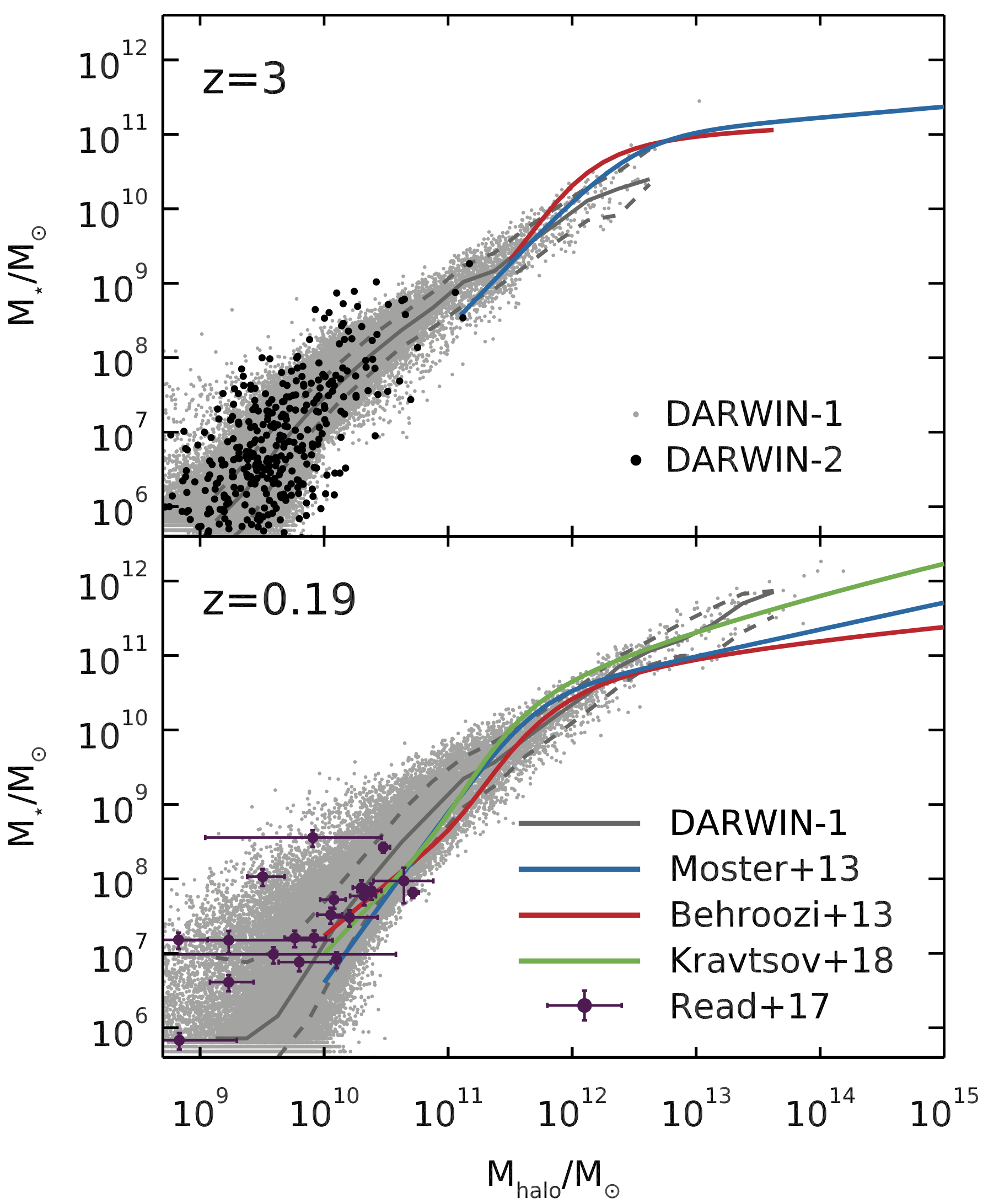}
\caption{Stellar mass of central galaxies as a function of halo mass taken to be $M_{\rm 200,c}$ in the local universe (bottom) and at $z=3$ (top). Small gray dots and large black dots show the distributions of the \darwini\ and \darwinii\ galaxies, respectively. The blue, red, and green solid lines indicate the semi-empirical relations from \citet{moster13}, \citet{behroozi13}, and \citet{kravtsov18}, respectively. Dark purple symbols with error bars show the relation for local dwarf galaxies \citep{read17}. The gray solid and dashed curves show the median and the $16^{\rm th}$--$84^{\rm th}$ percentiles, respectively, for the \darwini\ galaxies. In the local universe (bottom), \darwini\ shows moderate overall agreement with the empirical relations. At $z=3$, the \darwini\ and \darwinii\ galaxies are more massive than predicted by the empirical relations at $M_{\rm halo}\lesssim10^{11.5}\,\msun$, suggesting that the simulated galaxies grow earlier than expected in small halos.}
\label{fig:mstar_mhalo}
\end{figure}

The stellar mass--halo mass relation provides a measure of the efficiency with which baryons are converted into stars within dark matter halos. Stellar feedback is thought to regulate baryonic evolution in halos with $M_{\rm halo}\lesssim10^{12}\,\msun$, whereas AGN feedback becomes
increasingly important in more massive halos~\citep{silk98,bower06,croton08,somerville15}. 
Figure~\ref{fig:mstar_mhalo} shows the stellar mass--halo mass relation of \darwini\ at $z=3$ and $z=0.19$, together with the corresponding distribution for \darwinii\ galaxies at $z=3$. Here, $M_\star$ is the total stellar mass gravitationally bound to the central galaxy, as identified by \pgalf, while $M_{\rm halo}=M_{200,c}$ is the total mass enclosed within $R_{200,c}$. \darwinii\ follows a relation broadly similar to that of \darwini. At $z=3$, galaxies are generally more massive than the semi-empirical relation in halos with $M_{\rm halo}\sim10^{11.5}\msun$, implying overly efficient or premature stellar mass assembly at fixed halo mass. 

At lower redshift ($z=0.19$), the discrepancy between \darwini\ and the semi-empirical relations is reduced at $M_{\rm halo}\lesssim10^{11.5}\,\msun$. In particular, dwarf-scale galaxies in low-mass halos are broadly consistent with the local empirical relation of \citet{read17}. Around Milky-Way-mass halos, $M_{\rm halo}\sim10^{12}\,\msun$, the median relation is slightly below or comparable to the semi-empirical estimates. At higher halo masses, $M_{\rm halo}\gtrsim10^{12.5}\,\msun$, the \darwini\ relation lies above those of \citet{moster13} and \citet{behroozi13}, but is in good agreement with the relation of \citet{kravtsov18}.

This comparison highlights the sensitivity of the high-mass stellar mass--halo mass relation to the definition of galaxy stellar mass. In particular, the treatment of extended low-surface-brightness stellar components can substantially affect the inferred stellar masses of massive central galaxies. Therefore, the offset from the \citet{moster13} and \citet{behroozi13} relations should not be interpreted solely as evidence for insufficient quenching in \darwini. Nevertheless, the relatively high stellar masses of massive central galaxies remain sensitive to the efficiency of
late-time quenching and AGN feedback.

We note that these comparisons are not fully homogeneous. The relations of \citet{moster13}, \citet{behroozi13}, and \citet{kravtsov18} are derived using abundance matching for central galaxies, whereas \citet{read17} infer halo masses of dwarf galaxies from dynamical modeling of their rotation curves. Moreover, the inferred relation depends sensitively on the definition of stellar mass, particularly for massive galaxies with extended stellar envelopes. Differences in halo-mass definition, stellar-mass measurement, and the treatment of diffuse outer stellar components and satellites can therefore introduce systematic offsets among the different relations.

\begin{figure}
\centering 
\includegraphics[width=0.9\linewidth]{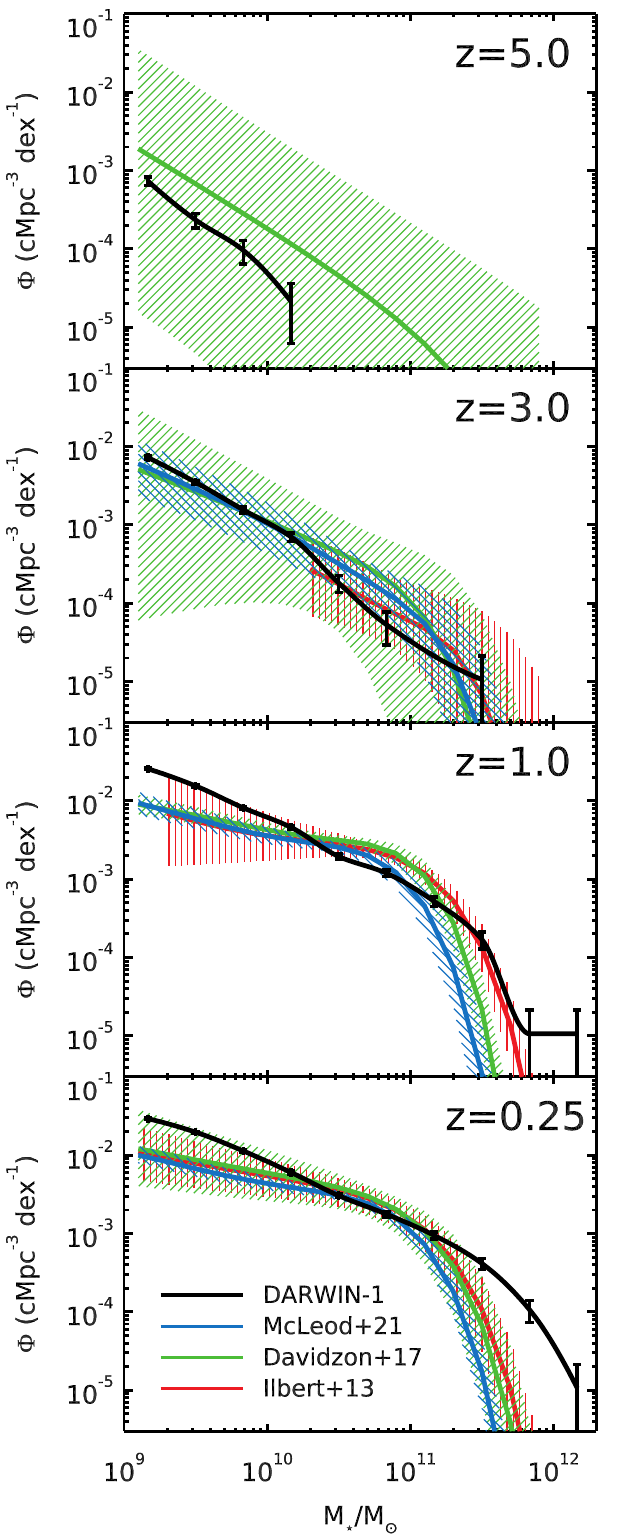}
\caption{Galaxy stellar mass functions at four different redshifts. The black solid curves show the \darwini\ GSMFs, while the blue, green, and red curves show the Schechter functions fitted by \citet{mcleod21}, \citet{davidzon17}, and \citet{ilbert13}, respectively. The black error bars indicate the Poisson errors of the \darwini\ GSMFs in individual bins. The hatched regions indicate the $1\sigma$ uncertainties obtained by Monte Carlo propagation of the Schechter-parameter errors presented in the literature.}
\label{fig:gsmf}
\end{figure}

\subsection{Galaxy stellar mass function of \darwini}\label{sub:gsmf}

The galaxy stellar mass function (GSMF) is a key statistic shaped by  the underlying physical processes that govern galaxy formation and evolution. Because the \darwinii\ simulations target only limited environments, we present the GSMFs of \darwini\ alone.
Figure~\ref{fig:gsmf} shows the GSMFs of \darwini\ together with three  observational estimates represented by Schechter-function fits. For the simulated GSMFs, we estimate the statistical uncertainty in each stellar mass bin from Poisson counting errors. The $1\sigma$ uncertainty range of the empirical GSMFs is estimated by Monte Carlo propagation of the asymmetric $1\sigma$ errors in the Schechter-function parameters given in the literature. At each stellar mass bin, the lower and upper $1\sigma$ bounds are defined as the $16^{\rm th}$ and $84^{\rm th}$ percentiles of the resulting distribution of the mass functions, assuming independent parameter uncertainties. Possible covariances among the Schechter parameters are therefore not included. Hence the hatched regions should be regarded as approximate  uncertainties rather than as complete observational error budgets.

\darwini\ shows moderate overall agreement with observations within 1$\sigma$ uncertainties at $z\gtrsim1$. At $z=5$, the simulated GSMF falls within the broad empirical uncertainty range, although the comparison is only weakly constraining because of the large uncertainties in the observed Schechter parameters and the limited number of massive galaxies in the simulation volume. 

As shown in Figure~\ref{fig:mstar_mhalo}, the stellar mass--halo mass relation of  \darwini\ at low redshift lies above those of \citet{moster13} and \citet{behroozi13} in halos with $M_{\rm halo}\gtrsim10^{12.5}\,\msun$, while remaining consistent with the relation of \citet{kravtsov18}. This spread among the semi-empirical relations indicates the sensitivity of the massive end to the definition and measurement of galaxy stellar mass, particularly to the treatment of extended low-surface-brightness stellar components.

The \darwini\ GSMF nevertheless shows an excess in the number density at $M_\star\gtrsim10^{11.5}\,\msun$ compared with the observational estimates. While this excess may partly indicate insufficient quenching and continued stellar mass growth in massive halos, its magnitude can also be affected by systematic differences in stellar-mass measurements. Such effects are particularly important at the exponentially declining massive end of the GSMF. In addition, the simulated GSMF in this regime is subject to substantial Poisson noise and cosmic variance because of the small number of massive galaxies.

The number density of galaxies with $M_\star<10^{10}\,\msun$ is also generally higher than observed, although it lies marginally within the $1\sigma$ uncertainty. This mild excess may indicate overly efficient star formation in low mass halos, but this regime is also more sensitive to numerical resolution and observational completeness. We therefore interpret the low-mass discrepancy with caution. In addition, as mentioned above, the uncertainty bands shown here are obtained by propagating the Schechter-parameter errors while assuming independent parameters.

\begin{figure}
\centering 
\includegraphics[width=0.975\linewidth]{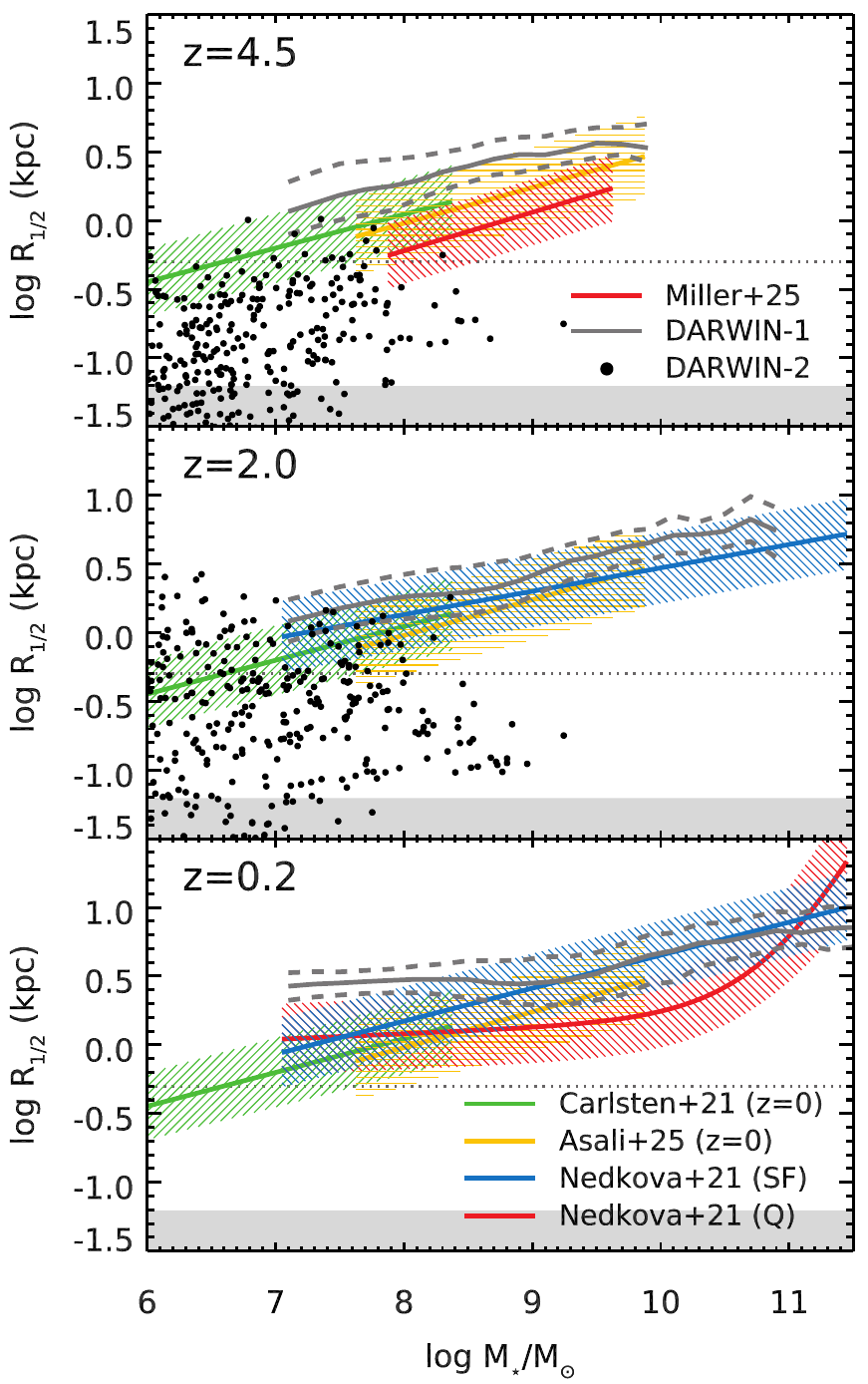}
\caption{Projected stellar half-mass radius $R_{1/2}$ as a function of total stellar mass at three redshifts. We measure $R_{1/2}$ in a two-dimensional projected plane. Black points show the $R_{1/2}$ values of the \darwinii\ galaxies at $z=4.5$ and $z=2.0$. Dark gray solid and dashed curves indicate the median and $16^{\rm th}–84^{\rm th}$ percentile ranges of the \darwini\ galaxies, respectively. In the top panel, the red curve shows the JWST mass–size relation at $z=4-5$ from \citet{miller25}. In the middle panel, the blue curve shows the fit for star-forming galaxies at $z=1.5-2.0$ from \citet{nedkova21}. In the bottom panel, the blue and red curves show the corresponding fits for star-forming and quiescent galaxies at $z=0.2-0.5$ from \citet{nedkova21}. Green and orange curves show the local dwarf-galaxy mass–size relations from \citet{carlsten21} and \citet{asali25}, respectively, plotted in all panels for comparison. Hatched regions denote the $1\sigma$ uncertainties obtained by Monte Carlo propagation of the fitting-parameter errors provided in the literature. Dark and light gray shaded regions mark the spatial-resolution limits of \darwini\ and \darwinii, respectively. At $z=4.5$, DARWIN-1 galaxies have larger $R_{1/2}$ than observed, but the offset decreases toward lower redshift. The \darwinii\ galaxies exhibit a wide range of sizes. }
\label{fig:size}
\end{figure}

\subsection{Galaxy mass--size relation}\label{sub:mass-size}

Figure~\ref{fig:size} shows the projected stellar half-mass radius ($R_{1/2}$) as a function of total stellar mass at three redshifts. The galaxies are projected onto the XY plane of the simulation volume, which effectively mimics a random viewing direction. At $z=4.5$, the \darwini\ galaxies are systematically larger than the observational mass--size relation of \citet{miller25}, with a typical offset of roughly 0.3 dex in $R_{1/2}$. The discrepancy decreases toward lower redshift, with the offset reduced to about 0.1 dex at $z=2$. In the low redshift snapshot, \darwini\ galaxies broadly follow the relation for star-forming galaxies from \citet{nedkova21}, although they remain larger than the local dwarf relations at low stellar masses. This low-redshift behavior is qualitatively consistent with the elevated low-redshift SFRD and the massive-end GSMF excess, suggesting that a substantial fraction of massive \darwini\ galaxies remains actively star-forming at late times. Continued in-situ star formation tends to build relatively compact central stellar structures, whereas subsequent dissipationless accretion and mergers can increase galaxy sizes by depositing stars at larger radii~\citep{naab09,lackner10,oser10,oser12,hilz13,dubois13}.

The larger sizes of low-mass \darwini\ galaxies are likely influenced by numerical resolution. Limited spatial resolution can impose an effective floor on galaxy sizes and provide artificial dynamical support to poorly resolved stellar systems \citep[e.g.,][]{dubois16,ludlow19}. Differences in mass resolution, star formation prescriptions, and feedback implementations may also contribute. In particular, bursty stellar feedback can drive rapid fluctuations in the gravitational potential and outward stellar migration, increasing galaxy sizes in low-mass systems \citep[e.g.,][]{teyssier13,el-badry16}.

The comparison between \darwini\ and \darwinii\ supports this interpretation. At $z=4.5$ and $z=2.0$, \darwinii\ galaxies are notably more compact than \darwini\ galaxies at $\log(M_\star/M_\odot)\simeq7$--9. In particular, many \darwinii\ dwarf galaxies lie below the empirical relation of star-forming galaxies at $z\simeq2$. However, this offset should be interpreted with caution. Compact objects with $R_{1/2}\sim100\,$pc are difficult to detect and resolve in high-redshift observations, and some of the smallest simulated systems approach the spatial-resolution limit.

Nevertheless, similarly compact dwarf galaxies are also found in the much higher-resolution \texttt{EDGE} simulations, which achieve spatial resolutions of a few parsecs and produce dwarf galaxies with half-light radii of order $\sim100$ pc while broadly reproducing the observed size distribution \citep{agertz20,rey25}. This suggests that such compact sizes are not necessarily numerical artifacts, although the \texttt{EDGE} galaxies are generally less massive and are primarily compared with observations at $z=0$. Moreover, the \texttt{EDGE} simulations show that dwarf-galaxy sizes can evolve substantially with assembly history, as late stellar accretion and dry mergers can deposit stars at large radii and produce much more extended systems \citep{rey19}.

In addition, \darwinii\ samples only a limited set of environments, including low-density field regions and the progenitor of a Milky-Way-like galaxy with its nearby filaments, whereas \darwini\ covers a much wider range of environments. Galaxy sizes can depend on environment, although both the magnitude and direction of this dependence vary with galaxy population and redshift \citep[e.g.,][]{maltby10,lani13,matharu19}. Recent JWST observations at $4.5<z<10$ find at most a weak environmental dependence of the mass--size relation~\citep{li25a}. Environmental selection may nevertheless contribute to part of the difference between the two simulations. 

Moreover, observed effective radii are luminosity-weighted, whereas our simulated sizes are stellar mass-weighted half-mass radii. Spatial variations in stellar populations can lead to either larger or smaller half-light radii relative to half-mass radii, depending on the distribution of young stars~\citep[e.g.,][]{suess19,ma18,degraaff22}. Dust can introduce an additional wavelength-dependent bias; in the rest-frame UV, centrally concentrated dust may preferentially suppress central star-forming emission and thereby increase the apparent galaxy size~\citep{shen24,kang26}.

Finally, the uncertainty ranges shown in Figure~\ref{fig:size} are obtained by propagating the published fitting-parameter errors while assuming that the parameters are independent. Possible covariances among the fitting parameters, uncertainties in size measurements, selection effects, and other observational systematics are not included. The comparison should therefore be regarded as a qualitative consistency check of galaxy sizes rather than a fully matched mock-observational analysis.

\subsection{Galaxy mass--metallicity relation}\label{sub:metallicity}

\begin{figure}
\centering 
\includegraphics[width=\linewidth]{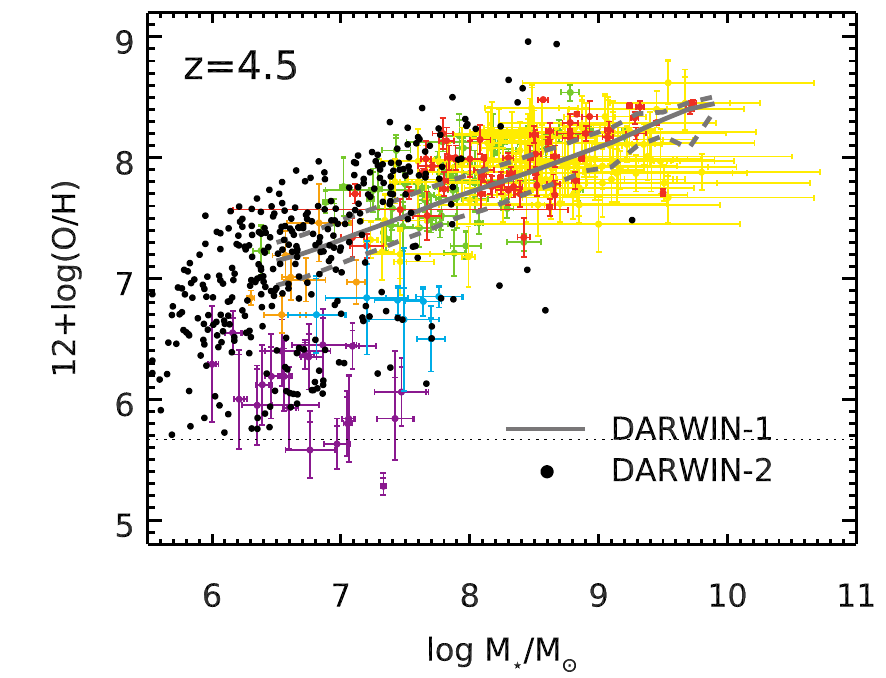}
\caption{Galaxy gas-phase metallicity as a function of stellar mass at high redshift ($z\sim4.5$). The gray solid and dashed lines denote the median and $16^{\rm th}-84^{\rm th}$ percentiles, respectively, for the \darwini\ galaxies, while the black dots indicate the relation for the \darwinii\ galaxies at $z=4.5$. The horizontal dash line indicates the metallicity floor of the simulations, $12+\log({\rm O/H})=5.67$. Colored symbols with error bars show observational measurements at $z=2.5-6.5$: \citet[][purple]{trussler26}, \citet[][yellow]{nakajima23}, \citet[][green]{curti24}, \citet[][red]{sarkar25}, \citet[][orange]{chemerynska24}, and \citet[][blue]{hsiao25}. \darwini\ is in good agreement with the observed trend at $12+\log {\rm O/H}>7$ while \darwinii\ reproduces the extremely metal-poor population at $\log M_\star/M_\odot<8$.}
\label{fig:oh_mgal}
\end{figure}

Figure~\ref{fig:oh_mgal} presents the galaxy gas-phase metallicity as a function of stellar mass at $z=4.5$. Since most of the observational data shown in Figure~\ref{fig:oh_mgal} are based on the JWST spectroscopy, we measure the oxygen and hydrogen abundances using all gas within a projected aperture of radius 2~proper kpc of each simulated galaxy center.
The slope and normalization of the \darwini\ galaxies are consistent with the observations. The \darwinii\ galaxies have overall slightly higher O/H than the \darwini\ galaxies at $\log(M_\star/M_\odot)>7$. 

A notable feature is the presence of extremely metal-poor galaxies (EMPGs) at $\log(M_\star/M_\odot)<8$ and $12+\log({\rm O/H})<6.7$ in the observational data, and \darwinii\ appears to reproduce this extremely metal-poor population. For reference, we adopt an initial metallicity of $10^{-3}\,Z_\odot$, with solar relative elemental abundances from \citet{asplund09}. This corresponds to a floor oxygen abundance of $12+\log({\rm O/H})=5.67$, as denoted by the horizontal dotted line in Figure~\ref{fig:oh_mgal}.

\citet{trussler26} infer from their stellar masses and star formation rates that the EMPGs are nearly pristine, young star-burst galaxies. Meanwhile, \citet{isobe26} suggest, based on JWST/NIRSpec spectra from the JADES full data release, that stochastic star formation in EMPGs can drive stellar feedback that expels metal-enriched gas, followed by accretion of more pristine material. Consistent with these interpretations, the EMPG analogs in \darwinii\ are chemically young systems with a single or episodic star formation histories. Their bursty star formation drives metal-enriched outflows, after which the galaxies gradually accrete more pristine gas. Their low oxygen abundances therefore indicate both limited prior enrichment and gas reservoirs dominated by pristine or weakly enriched material.

\begin{figure}
\centering 
\includegraphics[width=\linewidth]{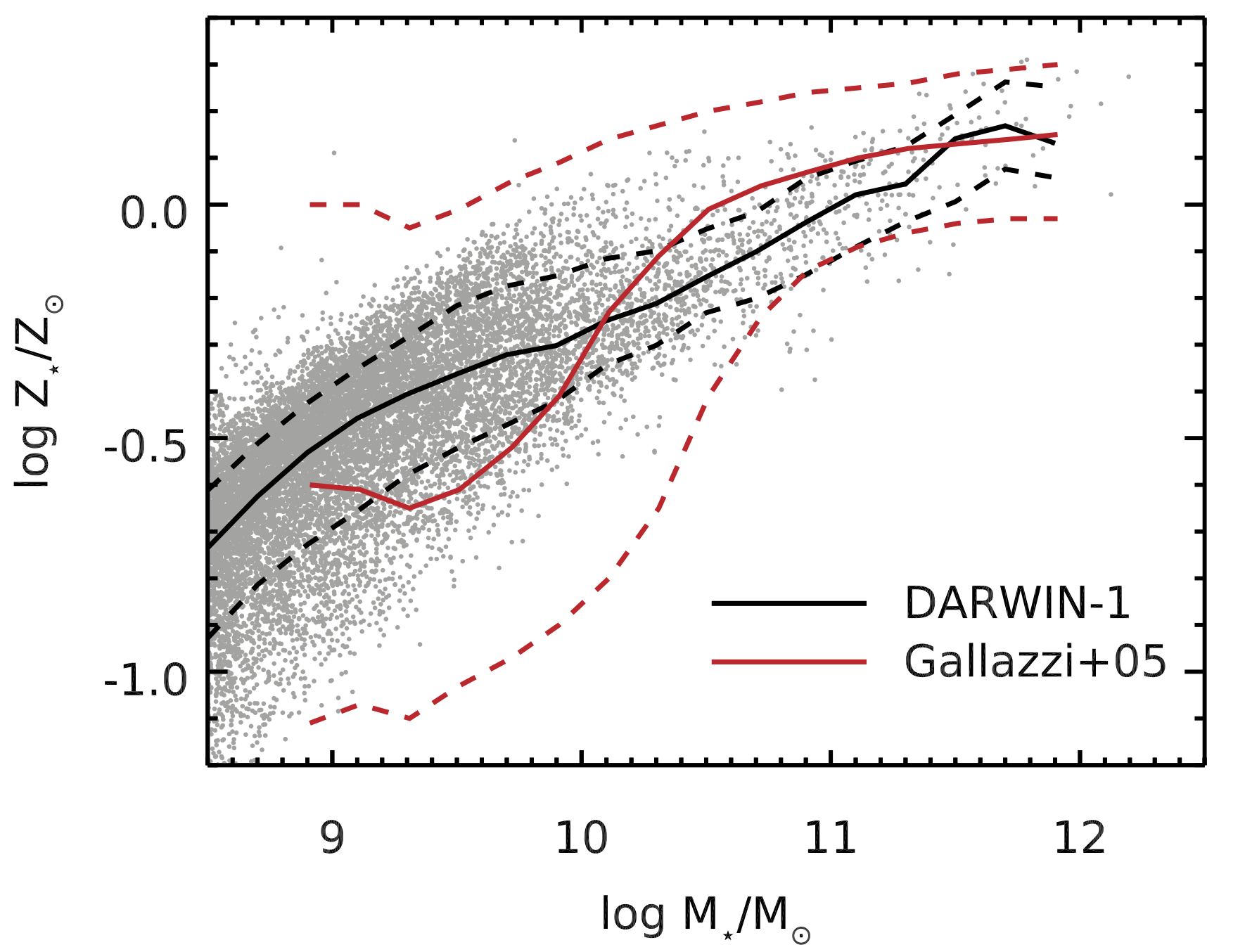}
\caption{Galaxy stellar metallicity as a function of stellar mass at $z=0.19$. Gray dots show individual galaxies while the black solid and dashed lines indicate the median and $16^{\rm th}-84^{\rm th}$ percentiles, respectively. The red curves show the observed relation in the local universe (median $z\sim0.13$) derived by \citet{gallazzi05} from the SDSS DR2 \citep{abazajian04}. \darwini\ shows moderate agreement with the observed relation within its 1$\sigma$ scatter, but has a different slope.}
\label{fig:z_mgal}
\end{figure}

At $z=4.5$, we identify 84 EMPGs in the entire \darwinii\ suite with stellar masses in the range $5.5<\log(M_\star/M_\odot)<8$, only two of which reside within the environmental influence of the Milky-Way-like progenitor, defined here as $3R_{200c}$. Consequently, the vast majority of these EMPGs are located in relatively low-density environments, such as the isolated field regions or filamentary structures connected to Milky Way-like galaxy. By targeting these underdense regions with \darwinii\ simulations, we are able to successfully reproduce and resolve the population of EMPGs. A comprehensive investigation into these EMPGs will be presented in future work.

Figure~\ref{fig:z_mgal} shows the distribution of the \darwini\ galaxies in the stellar mass--metallicity plane at $z=0.19$. Because the Sloan Digital Sky Survey Data Release 2 \citep[SDSS DR2;][]{abazajian04} uses 3-arcsec-diameter fibers, corresponding to a physical
diameter of approximately 7~kpc at its median redshift of $z=0.13$. We therefore measure the stellar metallicities of the simulated galaxies using the same aperture size. The \darwini\ galaxies show reasonable agreement with the observed relation within the $16^{\rm th}$--$84^{\rm th}$ percentile range, although the slope is shallower than observed at $\log(M_\star/M_\odot)<11$. The shallower slope of the simulated stellar mass--metallicity relation additionally indicates that the efficiency of baryon cycling is not uniform across halo mass. At low stellar masses, \darwini\ galaxies are slightly more metal-rich than the median trend of \citet{gallazzi05}, consistent with the mild excess of low-mass galaxies in the GSMF. In contrast, around $M_\star\sim10^{10.5}-10^{11}\,\msun$, the stellar metallicity is lower than observation, in line with the suppressed stellar mass growth in Milky-Way-mass halos seen in the stellar mass--halo mass relation. This suggests that star formation and metal recycling may be overly suppressed near the halo mass of the peak stellar mass fraction, while low mass systems may retain or produce metals too efficiently.


\begin{figure}
\centering 
\includegraphics[width=\linewidth]{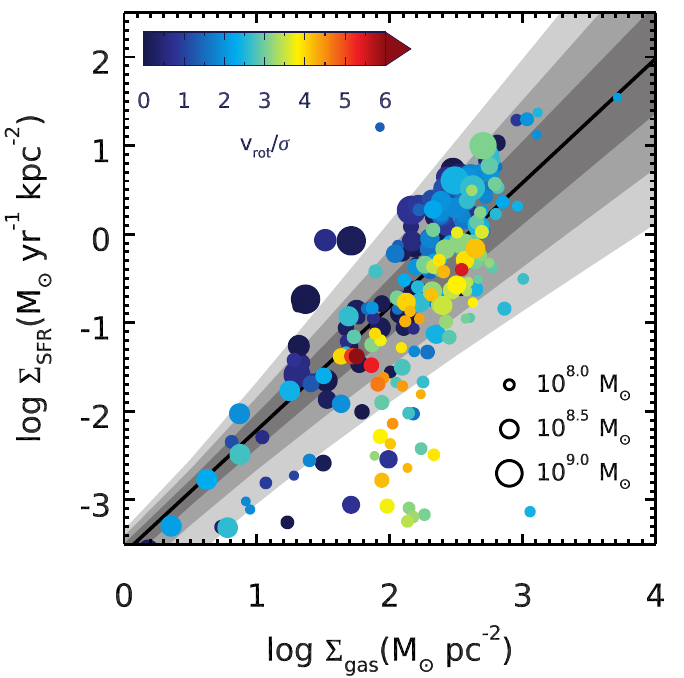}
\caption{Star formation rate surface density versus gas surface density for the \darwinii\ galaxies with stell mass $M_\star>10^8\,\msun$. The color coding denotes the ratio of rotation velocity to velocity dispersion of a gaseous disk, and the symbol size indicates stellar mass. The black solid line shows the observed relation derived by \citet{kennicutt98} for local spiral galaxies. The gray shaded regions indicate the $3\sigma$, $2\sigma$, and $1\sigma$ uncertainties of the empirical relation, from lighter to darker gray. The gas surface density is computed from the non-ionized hydrogen mass within $5R_{\star,\,1/2}$ and multiplied by a factor of 1.37 to account for helium and heavier elements, following \citet{kennicutt98}.The star formation rate surface density is derived from stars younger than 20~Myr within the same radius. Most galaxies lie within the $3\sigma$ range of the empirical relation, with galaxies having higher $V_{\rm rot}/\sigma$ tending to lie more closely along the Kennicutt relation. This trend suggests that the dynamical state of the gas disk may be linked to the regulation of star formation. The different timescales traced by gas and recent star formation may account for some outliers with high SFR surface density, whereas local thermo-turbulent conditions may lead to low SFR surface densities.}
\label{fig:sfr_gas}
\end{figure}

\subsection{Cold-gas surface density--star formation rate surface density relation}\label{sub:gas-sfr}

The SFR surface density ($\Sigma_{\rm SFR} $) is strongly correlated with the cold gas surface density~\citep[$\Sigma_{\rm gas}$,][]{kennicutt98}:
\begin{equation}
\begin{split}
\Sigma_{\rm SFR}=(2.5\pm0.7)\times 10^{-4} \\ 
\left( \frac{\Sigma_{\rm gas}}{ 1\msun\,{\rm pc^{-2}}} \right)^{1.4\pm 0.15}&\,\msun \, {\rm yr^{-1}\, kpc^{-2}}. 
\end{split}
\end{equation}
Because the ionization fractions of hydrogen and helium are explicitly tracked in \darwinii, we define the cold hydrogen gas mass as the non-ionized mass, $M_{\rm H}-M_{\rm HII}$, corresponding to $M_{\rm HI}+M_{\rm H_2}$. Following \citet{kennicutt98}, we multiply the hydrogen gas surface density by a factor of 1.37 to account for helium and heavier elements. 
Figure~\ref{fig:sfr_gas} shows this relation for the \darwinii\ galaxies with stellar masses greater than $10^{8}\,\msun$. The gas surface density is computed from the non-ionized hydrogen gas mass within $5R_{\star,\,1/2}$. The star formation rate is computed from the mass of stellar particles younger than 20~Myr within the same radius. Both surface densities are obtained by dividing the enclosed quantities by $\pi(5R_{\star,1/2})^2$. To characterize the gas kinematics, we define the disk plane to be perpendicular to the angular momentum vector of the gas within $5R_{\star,1/2}$. The rotation velocity, $V_{\rm rot}$, is the mass-weighted mean azimuthal velocity of the gas cells in this plane. We calculate the mass-weighted velocity dispersions in cylindrical coordinates, $\sigma_R$, $\sigma_\phi$, and $\sigma_z$, where the
$z$-axis is aligned with the gas angular momentum vector, and define
the total three-dimensional velocity dispersion as $\sigma=(\sigma_R^2+\sigma_\phi^2+\sigma_z^2)^{1/2}$. The color coding denotes the value of $V_{\rm rot}/\sigma$ for the gaseous component within $5R_{\star,\,1/2}$. 

The gray shaded regions indicate the $3\sigma$, $2\sigma$, and $1\sigma$ uncertainties estimated from the fitting errors. The shaded regions are obtained by propagating the fitting uncertainties of \citet{kennicutt98} and should not be interpreted as the full intrinsic scatter of the observed galaxy population. Most galaxies lie close to the empirical relation, typically within the range spanned by the propagated fitting uncertainty over the full ranges of redshift and stellar mass. Some galaxies fall below the $3\sigma$ range, possibly because local thermo-turbulent conditions lead to low formation efficiency.

Some of the scatter toward high SFR surface densities can be understood as a consequence of the different timescales traced by gas and star formation. While the gas surface density is measured instantaneously in a snapshot, the SFR surface density averages star formation over the past 20~Myr. After a burst, stellar feedback can rapidly reduce or redistribute the gas, whereas the SFR inferred from young stellar particles remains elevated for a short time. Such systems can therefore move temporarily above the time-averaged Kennicutt relation.

Interestingly, the scatter also appears to depend on the dynamical state of the gas. Galaxies with larger $V_{\rm rot}/\sigma$ tend to lie more closely along the Kennicutt relation, whereas dynamically hotter systems show a substantially broader distribution, including many of the high-SFR outliers. This suggests that the global relation between gas and star formation becomes more regular once the gas settles into a rotation-supported disk. For systems with low $V_{\rm rot}/\sigma$, our definition of gas surface density may also become less representative: the gas can be highly disturbed or spatially irregular, while $\Sigma_{\rm gas}$ is computed as a single disk-averaged quantity within $5R_{\star,\,1/2}$. Part of the increased scatter in these systems may therefore arise from both their non-equilibrium state and the adopted surface-density definition. Conversely, the galaxies on sequence may have reached some level of self-regulation, where star formation is regulated by their disc's dynamical response~\citep{park21}.

We find no strong systematic redshift dependence when the relation is
examined at different epochs. This suggests that, once a rotation-supported gas disk is established, the relation between gas and star formation surface densities is redshift independent over the range probed by \darwinii. Any cosmic evolution may therefore arise primarily through changes in the gas content and in the fraction of galaxies that have settled into rotation-supported disks, rather than through a strong evolution of the star formation relation itself~\citep{kraljic24}.

\begin{figure}
\centering 
\includegraphics[width=\linewidth]{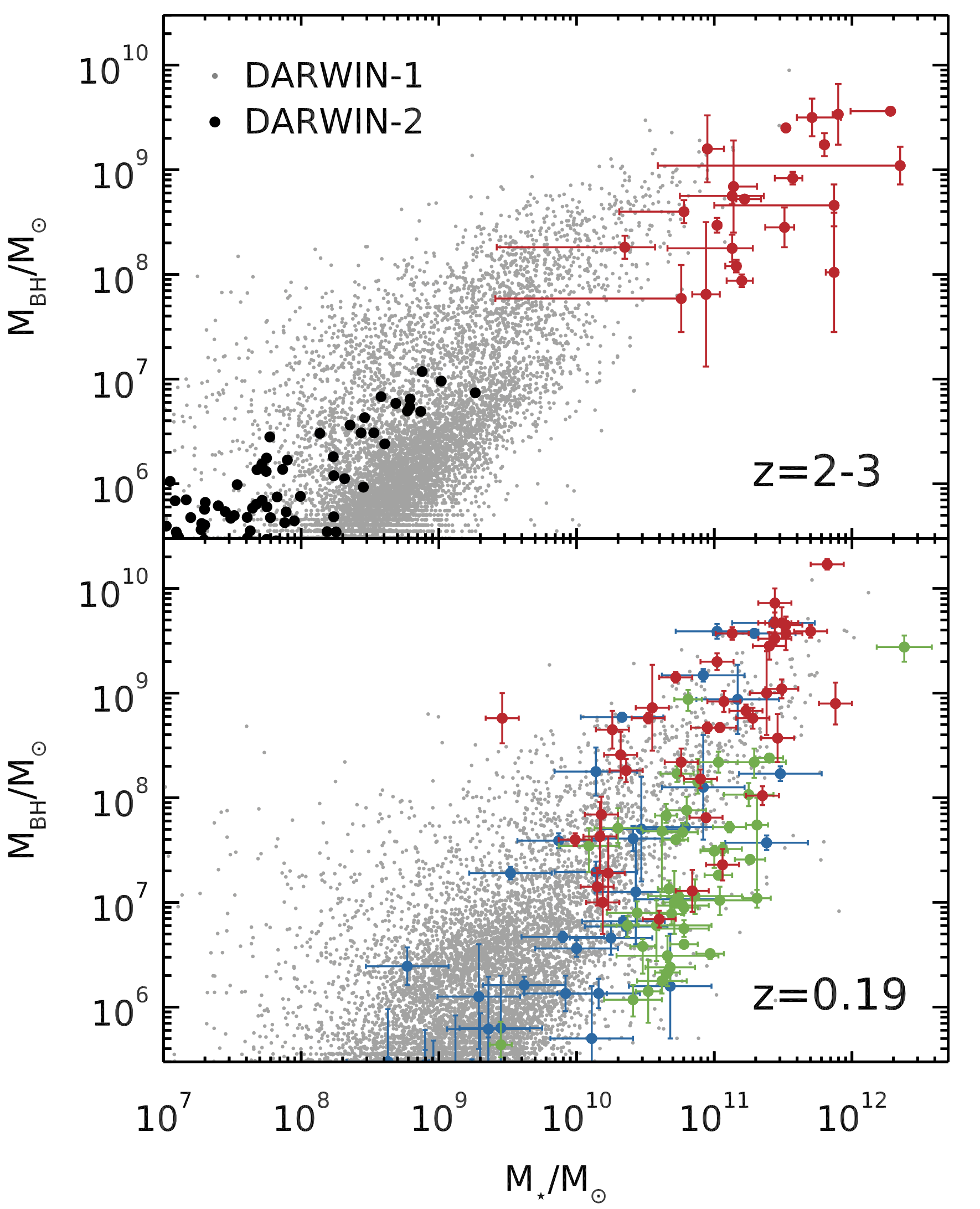}
\caption{Central black hole mass as a function of galaxy stellar mass in the local universe (bottom) and at $z=2$--$3$ (top). Small gray dots and large black dots show the distributions of the \darwini\ and \darwinii\ galaxies, respectively. In the top panel, the simulation data are taken from snapshots at $z=2.1$ for \darwini\ and $z\sim2-3$ for \darwinii\, while the red symbols with error bars indicate the observation of \citet{suh20} at $z>2$. In the bottom panel, the colored symbols show the observed relations from \citet[][blue]{reines15}, \citet[][green]{davis18}, and \citet[][red]{sahu19}. In \darwini, the relation agrees well with the observations in the local universe. At high redshift, however, the central black holes appear to grow ahead of the stellar mass assembly of their host galaxies, whereas the observed relation shows little evolution at $z<2$. }
\label{fig:mbh_mgal}
\end{figure}

\subsection{Black hole mass -- galaxy stellar mass relation}\label{sub:bh-stellar}

Figure~\ref{fig:mbh_mgal} shows the black hole mass--galaxy stellar mass relation at $z=2$--$3$ (top) and $z<0.2$ (bottom). In the local universe, \darwini\ reproduces the observed trend well. At $z=2$--$3$, however, the low-mass distributions of \darwini\ and \darwinii\ differ slightly. In \darwinii, central black holes are typically more massive than those in \darwini\ at fixed galaxy mass, even though the black hole seed mass in \darwini\ is twice as large as that in \darwinii. This difference is driven by more efficient gas accretion at higher resolution and by the smaller exclusion radius for black hole seeding, which leads to the formation of more black holes and, in turn, more frequent BH coalescences. This indicates that the high-redshift normalization of the relation is not set by the seed mass alone. We also find that the main locus at low galaxy masses represents the normal black hole population, whereas most objects scattered well above the relation are massive BHs hosted by satellite galaxies. These BHs likely experienced rapid growth in dense environments and were subsequently associated with relatively low-mass satellites through complex dynamical interactions. Although such systems are rare, they appear prominent because of their unusually high $M_{\rm BH}/M_\star$ ratios.

\begin{figure*}
\centering 
\includegraphics[width=\linewidth]{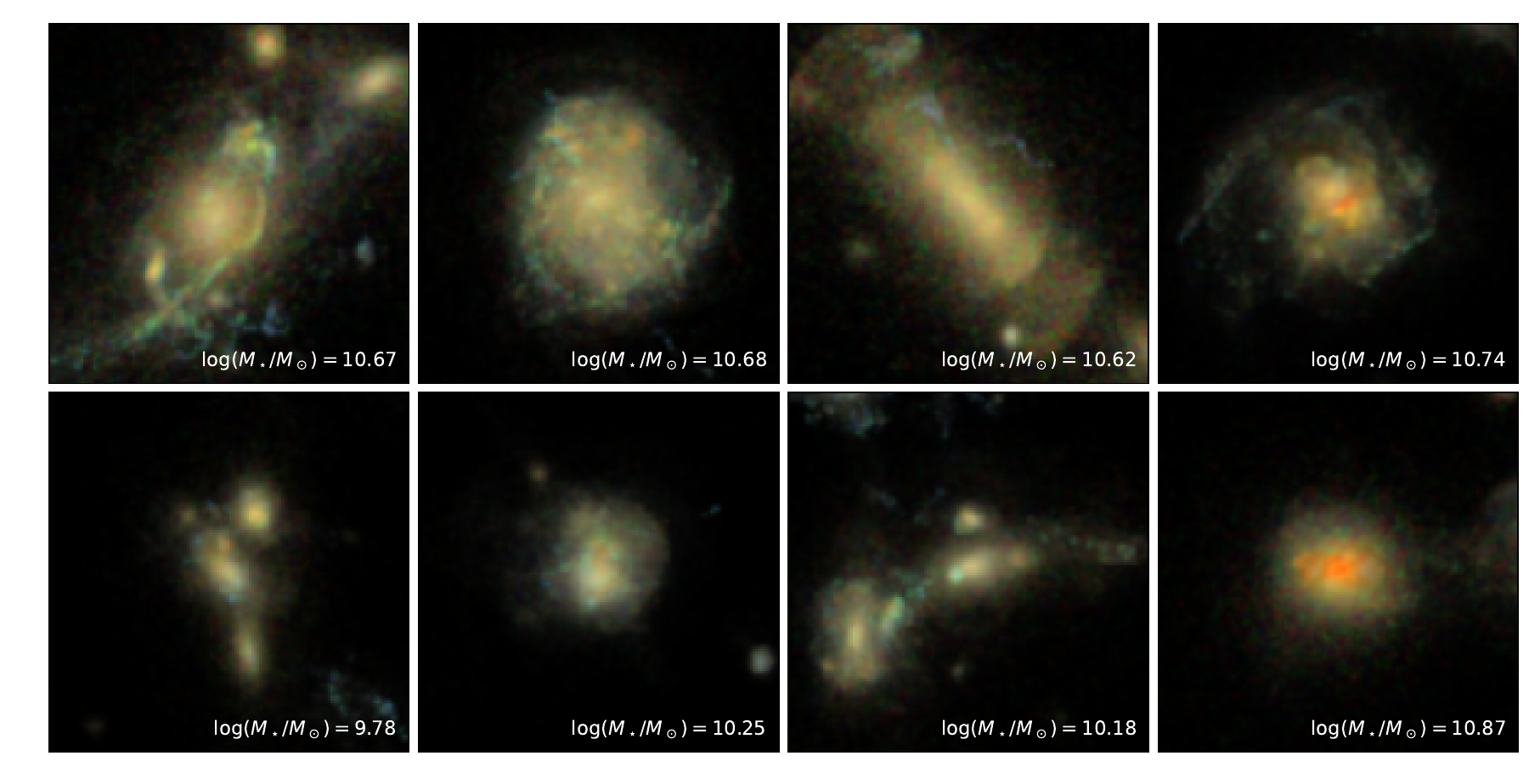}
\caption{Example of mock RGB images of \darwini\ galaxies at $z=1.76$ generated using \powderday. The composites are constructed from the observer-frame NIRCam F150W, F277W and F444W-band images, mapped to the blue, green and red channels, respectively, following the asinh-based RGB prescription of \citet{lupton2004}. Each panel spans a physical field of view of $80\, \rm kpc$ corresponding to 18.4 arc~sec in the cosmology adopted in \darwin. The stellar mass is indicated in the bottom right corner.} 
\label{fig:powderday}
\end{figure*}

\citet{suh20} argued, based on their results and previous studies in the local universe, that the black hole mass--galaxy mass relation evolves only weakly with time. In contrast, massive black holes in \darwini\ appear to grow earlier than the stellar assembly of their host galaxies. Indeed, we find that, on average, the stellar masses of galaxies with final stellar masses of $M_\star>10^7\,\msun$ grow about twice as fast as their central black holes over $z=3$--0.19. However, we note that black hole and galaxy mass estimates at high redshift may be affected by selection biases and AGN contamination~\citep[e.g.,][]{lauer07}. The observed systems in \citet{suh20} are luminous, broad-line AGN hosts, whereas the simulations include a much broader population of central black holes, including inactive or weakly accreting systems.


The apparent agreement in the local universe should be interpreted as a zeroth-order consistency check, because the local observational compilations combine different host galaxy selections, black hole mass measurements, and stellar mass definitions. For instance, \citet[][red]{sahu19} focus on early-type galaxies, whereas \citet[][green]{davis18} derive the relation for spiral galaxies. Both studies use black hole masses compiled from the literature that are directly measured from, for example, proper motions, stellar or gas dynamics, and/or astrophysical maser emission. By contrast, \citet[][blue]{reines15} estimate black hole masses for broad-line AGNs using single-epoch virial estimators. The comparison nevertheless indicates that the adopted black hole growth and feedback model produces a broadly realistic normalization at low redshift.

\section{Data Products}\label{sec:products} 
In this section, we present mock observations generated with the radiative transfer codes \skirt\ and \powderday\ in Section~\ref{sub:Mock}, together with a machine-learning application that uses higher-resolution \darwinii\ simulations to mitigate limitations induced by resolution in the full-box \darwini\ galaxy data in Section~\ref{sub:ML}. The \darwini\ simulation data and derived catalogs are publicly available through the \darwin\ data portal.\footnote{https://darwin.kasi.re.kr} Access to \darwinii\ data can be requested through a proposal-based process.

\subsection{Mock image generation} \label{sub:Mock} 



\begin{figure*}
\centering 
\includegraphics[width=\linewidth]{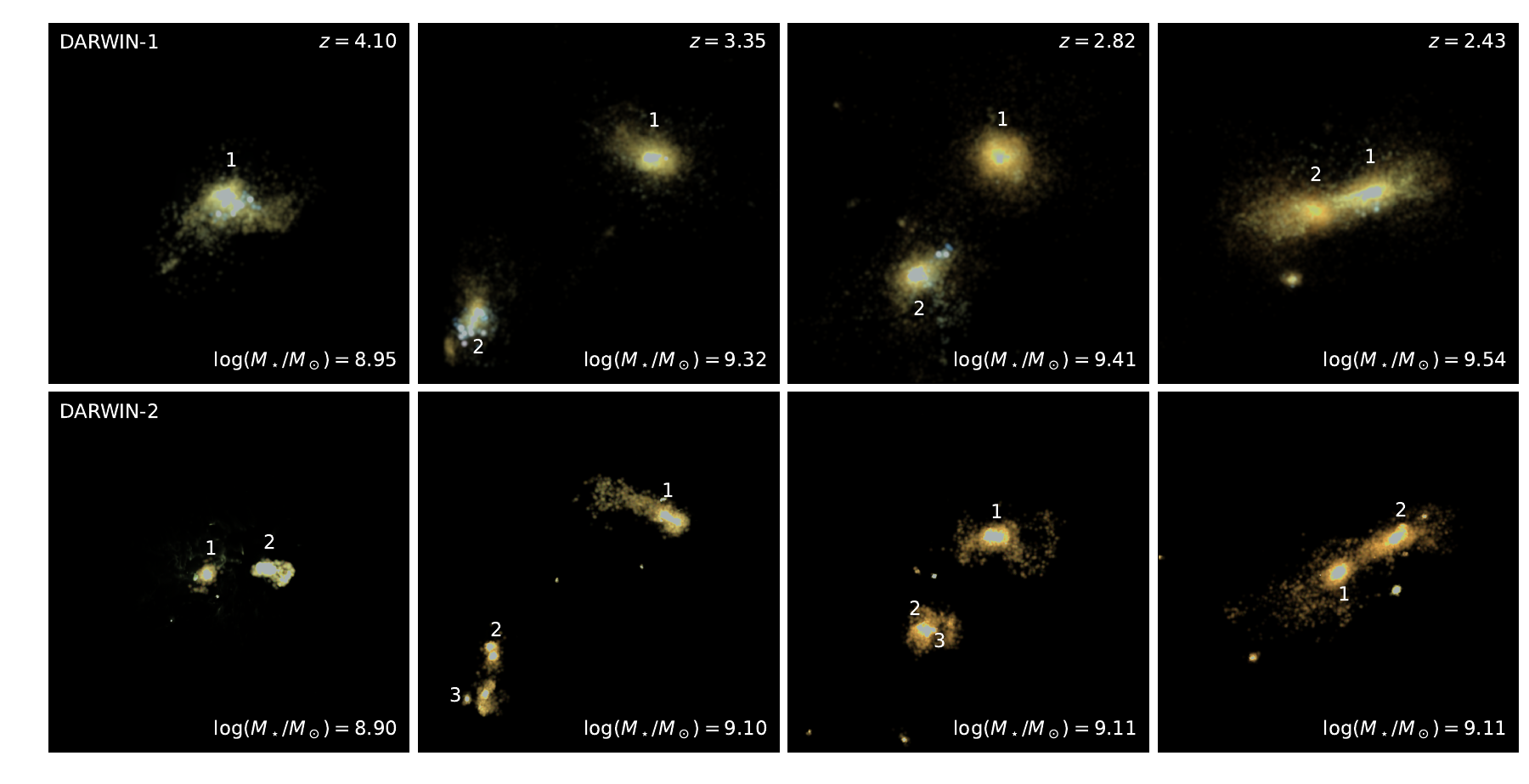}
\caption{Mock RGB images of matched galaxy systems in \darwini\ (top) and \darwinii\ (bottom) at four redshifts, from $z=4.10$ to $z=2.43$, generated using \skirt.
The composites are constructed from the rest-frame LSST $u$-, $g$-, and $r$-band images, mapped to the blue, green, and red channels respectively, following the asinh-based RGB prescription of \citet{lupton2004}.
Each panel spans a physical field of view of 100 kpc and is centered on the center of mass of the system.
The numbered labels identify galaxies with stellar masses greater than one-tenth of that of the most massive galaxy in each panel, ordered by decreasing stellar mass.
The stellar mass indicated in each panel is the total values summed over all galaxies within the field of view.
The physical pixel scales are 500 pc and 125 pc for \darwini\ and \darwinii, respectively.} 
\label{fig:darwin12_rbg_comp}
\end{figure*}

Synthetic observations provide an important bridge between simulated galaxies and observational data, allowing galaxy properties to be examined within an observationally motivated framework.
In particular, mock broadband images can be used to investigate the apparent morphology, colors, and surface-brightness distributions of simulated galaxies while accounting for the effects of stellar populations, dust attenuation, and viewing geometry.

As part of the \darwin\ project, we are constructing synthetic imaging products for simulated galaxies using radiative transfer postprocessing.
We employ the 3D Monte Carlo radiative transfer codes \skirt~ \citep{campsbaes2015,campsbaes2020} and \powderday~\citep{narayanan21}.
Our aim is not to perform a systematic comparison between the two radiative transfer pipelines, but rather to demonstrate the range of mock observation products that can be generated from the \darwin\ simulations.
A larger synthetic mock catalog is currently under construction for a variety of science applications, including machine-learning (ML) based matching between observed and simulated galaxies (Ha et al. in prep.).

For the radiative transfer calculations with \skirt\ and \powderday, stellar particles and gas cells are extracted from regions centered on selected galaxies and are used to construct the radiation sources and dusty interstellar medium, respectively.
In both calculation, the stellar emission is determined from the ages, masses, and metallicities of the stellar particles, while the dust distribution is constructed from the \ramses\ gas cells. We assume that a constant fraction of the metal mass, $f_{\rm dust}=0.4$, is locked into dust, consistent with values commonly adopted for galaxies in the local Universe \citep{dwek1998}. The detailed prescriptions for the stellar emission, young star-forming regions, dust properties, and radiative transfer differ between the two pipelines, as described below.

For the calculations with \skirt, stellar particles older than 1 Myr are assigned simple stellar population spectra from the Bruzual \& Charlot models, assuming a Chabrier IMF.
Stellar populations younger than 1 Myr are modeled separately using \ion{H}{2} region templates based on the MAPPINGS III library \citep{Groves2008} to account for emission associated with young star-forming regions. For the parameters describing the star-forming regions, including the ISM pressure, compactness, and PDR covering factor, are chosen following the prescriptions adopted for the TNG50-\skirt\ Atlas \citep{Baes2024}.
The spatial distribution of the stellar emission follows that of the stellar particles, with a smoothing length set by the distance to the 16${}^\mathrm{th}$ nearest stellar particle and capped at the minimum cell size of the each simulation; this scale also defines the pixel size of the \skirt\ images.
We adopt the THEMIS dust model \citep{Jones2017} to describe the wavelength-dependent absorption and scattering properties of the grains, and directly use the native adaptive-mesh structure of the \ramses\ simulations for the radiative transfer calculation.

For the calculations with \powderday, stellar spectra are generated using \texttt{FSPS} \citep{Conroy09, Conroy10} from the ages, masses and metallicities of the stellar particles, assuming a Kroupa IMF \citep{kroupa01}.
For stellar populations younger than 10 Myr, nebular emission is included using the precomputed CLOUDY-based lookup tables implemented in \texttt{FSPS} \citep{Byler17}.
While the same dust-to-metal ratio of $f_{\rm dust}=0.4$ is adopted, \powderday\ uses a Milky Way dust model with $R_V=3.1$, including the PAH component \citep{Draine03}.
The \ramses\ adaptive-mesh structure is reconstructed as a \texttt{yt} AMR dataset and passed to Hyperion \citep{Robitaille11}, which performs the Monte Carlo dust radiative transfer.


\begin{figure*}
  \centering
  \includegraphics[width=1.0\linewidth]{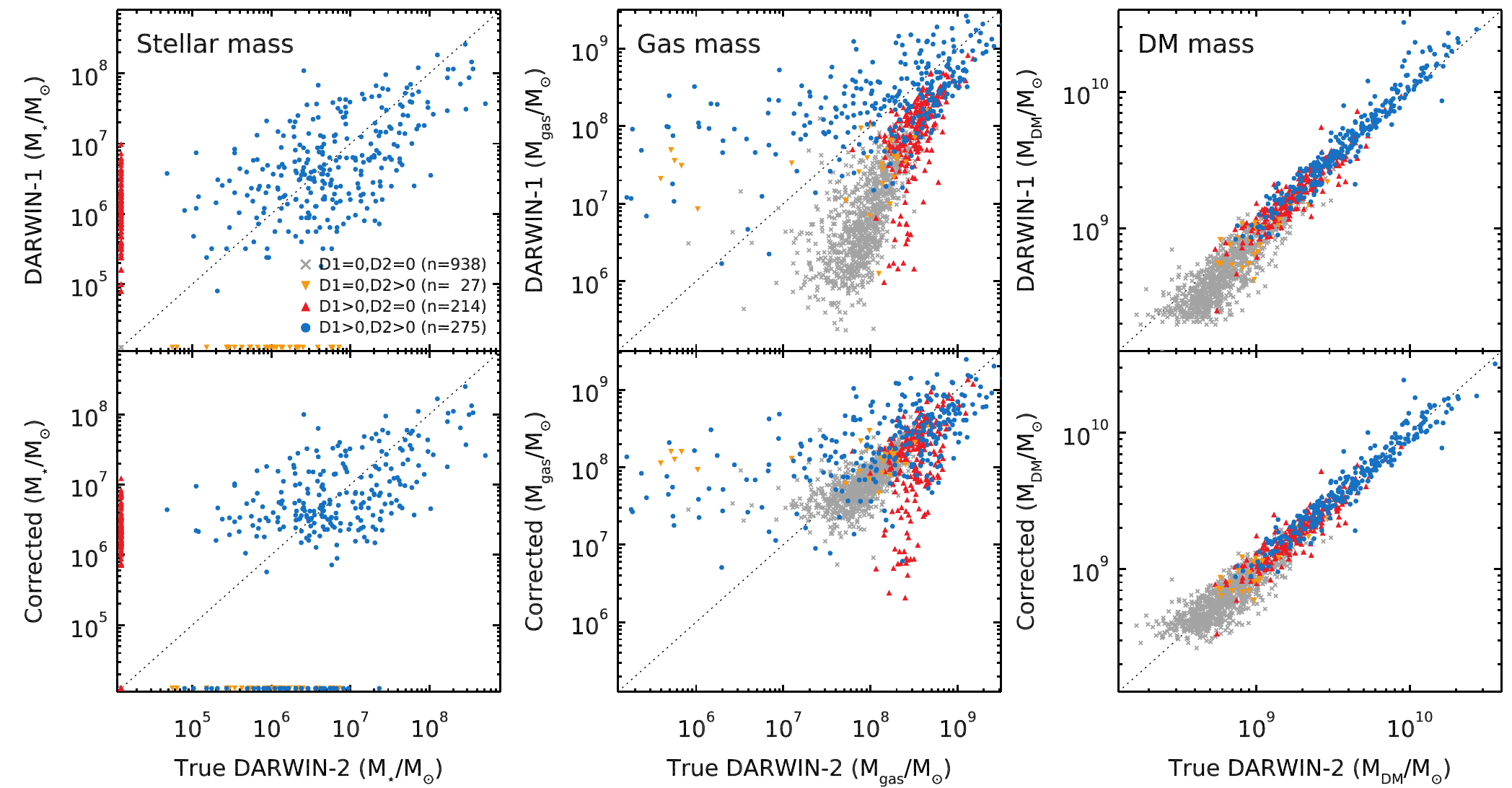}
  \caption{\textit{Top}: Comparison of matched \darwinii\ and \darwini\ halos in. stellar mass (left), gas mass (middle), and dark-matter mass (right) at $z=3.9$. Gas mass shows the strongest resolution-dependent offset, while dark-matter mass is much closer to convergence. \textit{Bottom}: Comparison between the matched \darwinii\ values and the ML-corrected \darwini\ predictions. The correction is most successful for gas mass, especially for systems without stars, while the stellar- and dark-matter-mass corrections show more modest improvements. The four symbol types represent halos with no stellar mass in either \darwini\ or \darwinii\ (gray crosses), stellar mass only in \darwinii\ (orange upside-down triangles), stellar mass only in \darwini\ (red triangles), and stellar mass in both \darwini\ and \darwinii\ (blue circles). }
  \label{fig:resolution-bias}
\end{figure*}

As one example, Figure~\ref{fig:powderday} presents mock JWST/NIRCam images generated with \powderday\ for eight \darwini\ galaxies at $z=1.76$ spanning a range of stellar masses and SFRs.
The observed frame F150W, F277W, and F444W-band images are mapped to the blue, green, and red channels, respectively, and combined using the asinh-based prescription of \citet{lupton2004}.
Each panel covers a physical region of $80\times80\,{\rm kpc}^2$.
The resulting images exhibit a wide variety of morphologies, including centrally concentrated systems, extended disks, asymmetric and interacting galaxies, and spatially resolved star-forming structures.
Despite their broadly comparable stellar masses, the galaxies span a large range in SFR and show substantial diversity in their resolved appearance.
These examples illustrate how \darwin\ galaxies can be transformed into observationally motivated imaging products suitable for comparison with JWST observations.

We use \skirt\ for a complementary example that follows the evolution of a matched galaxy system in \darwini\ and \darwinii.
Figure~\ref{fig:darwin12_rbg_comp} shows the system at four epochs, from $z=4.10$ to $z=2.43$.
The RGB composites are constructed from rest-frame LSST $u$-, $g$- and $r$-band images, mapped to the blue, green, and red channels, respectively.
Each panel covers a physical region of $100\times100\,{\rm kpc}^2$.
The physical pixel scales are 500 pc for \darwini\ and 125 pc for \darwinii, reflecting the different spatial resolutions of the two simulations.
We launch $10^6$ and $1.6\times10^7$ photon packets for \darwini\ and \darwinii, respectively, which we find to provide adequate convergence for the broadband images considered here.

The matched sequence illustrates how differences in numerical resolution and star-formation history are reflected in synthetic observations. Although the two systems correspond to the same large-scale environment and have similar total stellar masses, their star-formation histories differ substantially. \darwinii\ forms most of its stars earlier and shows little ongoing star formation at later epochs, resulting in redder stellar populations and a lack of the prominent blue star-forming regions seen in \darwini, which retains more recent star formation and more extended blue stellar components. At the same time, the finer spatial resolution of \darwinii\ allows compact stellar structures and smaller satellite systems to be resolved more clearly. The overall color difference therefore appears to be driven primarily by the different star-formation histories, whereas the differences in resolved small-scale structure are more closely associated with the different spatial resolutions.

\begin{deluxetable}{lcccc}
\tablecaption{Summary performance comparison between the native \darwini\ values, the halo-mass-only correction, and the full-feature ML correction.}
\label{tab:ml-performance}
\tablehead{Quantity & Metric & Native resolution bias & Halo-mass-only correction & ML correction}
\startdata
Stellar occupancy & ROC-AUC\tablenotemark{a} & 0.866 & 0.928 & 0.934 \\
Stellar mass & RMSE [dex]\tablenotemark{b} & 0.634 & 0.625 & 0.629 \\
Gas mass & RMSE [dex] & 1.060 & 0.592 & 0.515 \\
DM mass & RMSE [dex] & 0.208 & 0.170 & 0.169
\enddata
\tablenotetext{a}{Area under the receiver operating characteristic curve (ROC-AUC). For the native \darwini\ stellar occupancy, the reported ROC-AUC is computed from binarized predictions and is equivalent to the average of the true positive rate and true negative rate.}
\tablenotetext{b}{Root-mean-square error. For stellar mass, the quoted RMSE is computed for the subset with $M_{\star,\darwini}>0$, $M_{\star,\darwinii}>0$, and $M_{\star,\rm prediction}>0$.}
\end{deluxetable}

Taken together, these examples demonstrate the flexibility of the \darwin\ mock observation framework.
Both \powderday\ and \skirt\ can be used to generate radiative-transfer-based synthetic data products that can subsequently be converted into broadband images tailored to specific observational facilities and filter sets, including JWST-like imaging.
Here, we simply present two representative applications: a \powderday-based montage of \darwini\ galaxies and a \skirt-based comparison of a matched galaxy system in \darwini\ and \darwinii.
These examples are intended to illustrate the range of synthetic imaging products that can be constructed from the \darwin\ simulations, rather than to provide a quantitative comparison between the two radiative transfer codes or a statistical analysis of galaxy morphology.



\subsection{Resolution bias and ML 
Applications} \label{sub:ML} 



Explicit resolution convergence tests have been carried out to assess the predictive power of cosmological hydrodynamical simulations \citep{2015MNRAS.450.1937C}. In parallel, ML approaches have increasingly been used to learn mappings between lower-cost simulation outputs and higher-fidelity simulations \citep[e.g.,][]{2019MNRAS.489.3565J,2022MNRAS.509.5046L,2026MNRAS.545f2096D}. Most directly relevant to the present work, \citet{2024ApJ...965..156J} used matched galaxies across IllustrisTNG simulations at different resolutions, assuming that the same galaxies are present in both runs. They showed that baryonic quantities can exhibit systematic cross-resolution offsets, and that these offsets can be corrected using a tree-based regression model. Based on this framework, we use ML here as a proof-of-concept diagnostic of which \darwini\ properties admit a transferable correction to their matched \darwinii\ counterparts. A detailed description of the methodology is provided in Appendix \ref{sec:appendix-ml}. Our matched sample contains 1454 halos across eight different \darwinii\ zoom-in fields at $z\sim3.9$, including 275 galaxies with $M_{\star}>0$ in both simulations.  

\begin{figure*}
  \centering
  \includegraphics[width=1.0\linewidth]{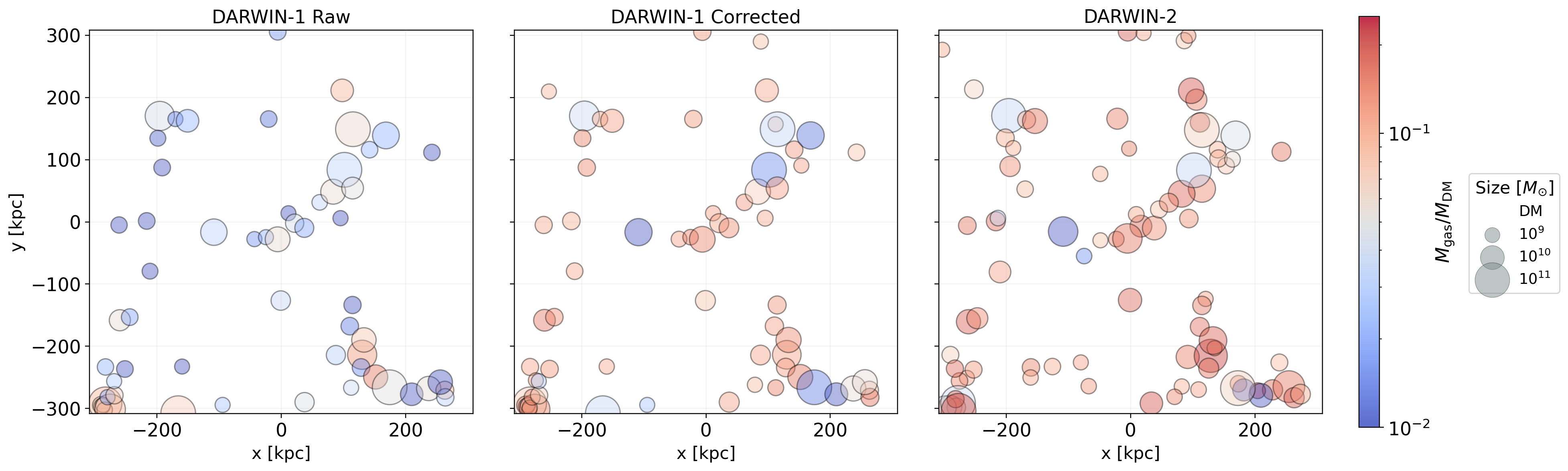}
  \caption{Example of ML correction of galaxy properties. Circles represent dark matter halos, color-coded by the gas fraction $M_{\rm gas}/M_{\rm DM}$. The left panel shows the distribution of galaxies in the MW-filament zoom-in region in \darwini, the middle panel shows galaxies with physical quantities corrected by the ML model, and the right panel shows the results from \darwinii. }
  \label{fig:ml-snap}
  \vspace{2mm}
\end{figure*}

The intrinsic offset pattern between \darwini\ and \darwinii\ is displayed in the upper panels of Figure \ref{fig:resolution-bias}. The strongest systematic resolution bias is found in the gas mass, with a median offset of $\sim 0.82$ dex, such that galaxies in \darwinii\ are generally more gas-rich than their matched counterparts in \darwini. This offset becomes increasingly severe toward the low-mass end. In contrast, for the subset of galaxies with $M_{\star}>0$ (blue circles), the gas-mass offset is much weaker and can even be slightly reversed, with \darwini\ showing comparable or modestly higher gas masses. This may indicate that, once star formation is present, the lower-resolution run tends to retain somewhat more gas, possibly reflecting less efficient stellar-feedback-driven gas removal or differences in the sampling of star formation and feedback at low resolution. By contrast, the dark-matter mass is already close to convergence, with a median offset of only $\sim 0.06$ dex, and therefore serves as a useful control quantity. Stellar mass shows the largest scatter, with no clear systematic bias among the matched luminous systems. Still, we note that a significant fraction of low-mass dwarfs have $M_{\star} > 0$ only in \darwini. This difference in stellar occupancy is also a target of the ML-based correction.

As a baseline nonlinear model, we adopt LightGBM \citep{10.5555/3294996.3295074} and train it to predict the matched \darwinii\ properties from the \darwini\ inputs, i.e., to learn a low-to-high-resolution correction. All quoted test set results are obtained from leave-one-field-out (\texttt{GroupKFold}) validation across the different zoom-in regions; that is, halos in one zoom-in region are predicted using a model trained on the other regions. Thus, the reported performance reflects \textit{transfer} across independent zoom-in regions rather than random-split samples. The lower panels of Figure \ref{fig:resolution-bias} compare the \darwinii\ values with the leave-one-field-out ML predictions in a test set. The correction is most successful for gas mass, where the predicted points move substantially closer to the one-to-one relation. However, it does not properly capture the bimodality of the gas distribution mentioned above, exhibiting poorer performance in correcting gas mass for halos with $M_{\star}>0$. This suggests that the gas-resolution bias is not well described by a single global mapping. The stellar-mass correction is primarily limited by occupancy: the main improvement arises from correctly identifying whether a halo is luminous in \darwinii, rather than correcting the stellar mass of dwarf galaxies with $M_{\star} > 0$. The improvement for dark-matter mass is modest, as expected for a quantity that is already close to numerical convergence.

To assess the utility of a multivariate ML correction, we compare three predictors in Table \ref{tab:ml-performance}: the native \darwini\ values, a halo-mass-only model that uses only $M_{\rm dm,\darwini}$, and a full-feature model that uses the full \darwini\ feature set. The halo-mass-only model represents a low-dimensional statistical correction for resolution bias, which is one way to mitigate resolution bias \citep[e.g.,][]{2018MNRAS.475..648P}. The full ML model tests whether additional baryonic, structural, or environmental information carries predictive signal beyond halo mass alone. Stellar occupancy is evaluated with ROC-AUC, where higher values indicate better discrimination between \darwinii-luminous and \darwinii-dark halos. The remaining rows report RMSE in log space, where lower values indicate better recovery of the matched \darwinii\ quantities.

The strongest improvements occur for gas mass and stellar occupancy. Gas mass improves substantially relative to both the native \darwini\ values and the halo-mass-only correction, indicating that additional \darwini\ properties contain useful information beyond a simple halo-mass-dependent shift. Stellar occupancy also improves over the native \darwini\ luminous indicator, although the gain over the halo-mass-only model is modest. In contrast, the conditional stellar-mass regression and the dark-matter-mass correction show little or no improvement over the halo-mass-only baseline. This indicates that stellar occupancy is largely predictable from halo mass, with only a modest additional gain from the full-feature ML model, while dark matter is already close to numerical convergence.

Overall, this exercise provides a proof-of-concept assessment of resolution bias rather than a production-ready correction scheme. The robust conclusion is not that the ML model used in this work is the optimal algorithm, but rather that ML can effectively correct physical properties, albeit with limited performance in stellar mass due to the small dataset size. Because the models are trained on data from different environments and evaluated across them, the results provide an initial test of transfer across independent zoom-in regions, but the modest size of the matched sample---especially the few hundred luminous galaxies---limits the strength of the conclusions. Figure \ref{fig:ml-snap} illustrates one example of the corrected halo catalog, showing that the ML correction improves the low-mass gas content but still underpredicts some of the gas-rich and clustered structure present in \darwinii. Future work should therefore focus on uncertainty-aware corrections, improved treatment of luminous massive dwarfs, and physically motivated low-dimensional models that can be calibrated more robustly with limited matched samples.

\section{Summary}
\label{sec:summary}

The \darwin\ project is designed to study the formation and evolution of dwarf galaxies in a cosmological context through a multi-level simulation strategy using the cosmological hydrodynamical code \ramses. To sample a wide range of environments, we perform the full-box simulation \darwini\ in a volume of $(65.536\,{\rm cMpc})^3$, with a minimum cell size of $\Delta x=500\,{\rm pc}$ and mass resolutions of $M_{\rm DM}=8.9\times10^6\,\msun$ and $M_{\star}=10^5\,\msun$. 

We also perform the \darwinii\ zoom-in simulations to resolve dwarf galaxies on sub-100-pc scales while retaining their cosmological environments.
We select 24 field regions and a Milky Way analog together with its connected filaments from the \darwini\ DM-only run and resimulate them at a finer spatial resolution of 62.5 pc, with mass resolutions of $M_{\rm DM}=1.7\times10^4\,\msun$ and $M_{\star}=10^4\,\msun$. The baryonic physics is tailored to the resolution of each simulation. In \darwini, star formation is modeled using a simple density threshold, whereas \darwinii\ adopts a thermo-turbulent prescription based on the local gas properties. In addition, radiative transfer is performed in three photon energy bands, allowing the non-equilibrium ionization states of hydrogen and helium to be followed explicitly.

To fully utilize high-performance computing resources based on many-core architectures, we implement OpenMP parallelism in \ramsesrt\ on top of its MPI parallelization. The Godunov and radiative-transfer routines show good scaling because of their high computational locality. We identify galaxies and subhalos from heterogeneous matter components using \pgalf~\citep{kim23}, and construct merger trees using stellar-particle IDs for galaxies and dark-matter-particle IDs for subhalos. 

We compared the properties and scaling relations of \darwini\ and \darwinii\ galaxies with observations to assess the performance of the simulations. To facilitate a more direct comparison between simulated galaxies and observational data, we also generate mock images using the radiative transfer codes \skirt\ and \powderday. In numerical experiments, resolution, together with the adopted physical prescriptions, can strongly influence the properties of simulated galaxies. To mitigate resolution-driven biases between \darwini\ and \darwinii, we adopt a proof-of-concept approach based on a machine-learning technique. These results can be summarized as follows.

\begin{itemize}

\item \darwini\ reasonably reproduces the evolution of the cosmic star formation rate density and the galaxy stellar mass functions at $z>1$. At lower redshift, the stellar mass--halo mass relation is broadly consistent with empirical constraints, while the galaxy stellar mass function shows an excess at $M_{\star}>10^{11.5}\,\msun$. This may partly reflect insufficient late-time quenching in massive systems, although the inferred stellar masses of massive galaxies are sensitive to their adopted definition and measurement.

\item The stellar half-mass radii ($R_{1/2}$) of \darwini\ galaxies are systematically larger than observed, with a typical offset of roughly 0.3 dex at $z=4.5$ that decreases to about 0.1 dex at $z=2$. This may partly reflect the relatively coarse spatial and mass resolution and differences in the stellar-feedback implementation in \darwini. In \darwinii, dwarf galaxies show a wide range of $R_{1/2}$, including compact systems that are difficult to resolve observationally at high redshift.

\item The gas-phase metallicities of galaxies in \darwini\ and \darwinii\ are broadly consistent with recent JWST observations at high redshift. In particular, \darwinii\ reproduces extremely metal-poor galaxy candidates with $12+\log({\rm O/H})<6.7$ at $M_\star<10^{8}\,\msun$~\citep[see][]{trussler26}. Our preliminary analysis suggests that the EMPG analogs in \darwinii\ are chemically young systems formed by bursty
star formation, metal-enriched outflows, and subsequent pristine gas accretion. The stellar metallicities of \darwini\ galaxies are also in reasonable agreement with local observational constraints~\citep{gallazzi05}, although the simulated relation has a shallower slope.

\item We examined the relation between star formation rate surface density and gas surface density in \darwinii. Galaxies with rotation-dominated gas kinematics broadly follow the Kennicutt relation for local spiral galaxies~\citep{kennicutt98}, with systems having higher $V_{\rm rot}/\sigma$ tending to lie more tightly along the empirical relation. This suggests that the dynamical settling of the gas diskis associated with the emergence of a tighter gas--star formation relation

\item In \darwini, massive black holes tend to assemble earlier than the
stellar components of their host galaxies, whereas recent high-redshift observations suggest little evolution in the black hole mass--galaxy stellar mass relation~\citep[see][]{suh20}. This comparison should be interpreted with caution because high-redshift black hole and host-galaxy mass estimates can be affected by selection biases and AGN contamination. At $z<0.2$, \darwini\ shows good agreement with observations in the $M_{\rm BH}$--$M_\star$ plane.

\item We generate mock images of selected \darwini\ and \darwinii\ galaxies using the 3D dust radiative transfer codes \skirt\ and \powderday. The mock images reveal a diversity of galaxy morphologies and demonstrate how differences in spatial resolution and star formation histories between the two simulations are reflected in their observed appearance.

\item We apply a proof-of-concept ML model calibrated on matched \darwini--\darwinii\ halos. The model substantially improves gas-mass predictions and stellar occupancy, whereas corrections to stellar and dark matter masses are modest.

\end{itemize}

Although the \darwin\ simulations do not capture all physical processes relevant to galaxy formation, they provide an independent theoretical data set that complements existing observations and offers a useful reference for examining the environmental dependence of dwarf galaxy formation. Toward this broader goal, \darwiniii\ is being developed to extend the \darwin\ suite to the internal scales of individual dwarf galaxies.
The \darwini\ simulation data and derived catalogs are publicly available through the \darwin\ data portal\footnote{https://darwin.kasi.re.kr}. Access to \darwinii\ data is provided through a proposal-based process.

\section*{acknowledgments}
This work is supported by the National Research Foundation (NRF) of Korea grant funded by the Korea government (MSIT, RS2022-NR068800). TK was supported by the National Research Foundation of Korea (RS-2022-NR070872 and RS-2025-00516961), and by the Yonsei Fellowship, funded by Lee Youn Jae. J.-H.K.’s work was supported by the National Research Foundation of Korea (NRF) grant funded by the Korea government (MSIT) (No. 2022M3K3A1093827 and No. 2023R1A2C1003244).  His work was also supported by the Global-LAMP Program of the NRF funded by the Ministry of Education (No. RS-2023-00301976). This work benefited from the outstanding support provided by the KISTI National Supercomputing Center and its Nurion Supercomputer through the Grand Challenge Program (KSC-2024-CHA-0006, KSC-2024-CHA-0025, KSC-2024-CHA-0028, KSC-2024-CRE-0478). Large data transfer was supported by KREONET, which is managed and operated by KISTI. This work is also supported by the Center for Advanced Computation at Korea Institute for Advanced Study. This work is partially supported by the grants GALBAR ANR-25-CE31-4684.
\appendix
\label{sec:appendix}
\restartappendixnumbering

\section{Summary of the \darwinii\ zoom-in simulations}
\label{sec:sim_list}

\begin{deluxetable}{cccccc}
\tablecaption{List of the \darwinii\ simulations}
\label{tab:sim_list}
\tablehead{
\colhead{Run} & \colhead{Region} & \colhead{\shortstack{Volume}} & \colhead{\shortstack{Most massive halo}} & \colhead{Latest snapshot} & \colhead{Tracked elements}\\[-1ex]
\colhead{} & \colhead{} & \colhead{[$\rm cMpc^3$]} & \colhead{[$M_{\rm 200c}/\msun$]} & \colhead{[$z$]} & \colhead{}
}
\startdata
01 &MW-filament & 19 & $2.4\times10^{12}$ & 2.12 &  H, N, O, Mg, Fe\\
\hline
02&Field 01 & 64 & $1.1\times10^{11}$ & 0.00  & H, N, O, Mg, Fe \\
\hline
\multirow{4}{*}{03}
&Field 02 & 64 & $3.1\times10^{11}$ & \multirow{4}{*}{2.71} & \multirow{4}{*}{H, O, Mg, Fe}  \\
&Field 03 & 64 & $2.9\times10^{11}$ &  &  \\
&Field 04 & 64 & $7.2\times10^{10}$ &  &  \\
&Field 05 & 64 & $7.2\times10^{10}$ &  &  \\
\hline
\multirow{4}{*}{04}
&Field 06 & 64  & $1.4\times10^{11}$ & \multirow{4}{*}{2.52} & \multirow{4}{*}{H, O, Mg, Fe} \\
&Field 07 & 64 & $5.8\times10^{10}$ &  &  \\
&Field 08 & 64 & $1.1\times10^{11}$ &  &  \\
&Field 09 & 64 & $6.1\times10^{10}$ &  &  \\
\hline
\multirow{3}{*}{05}
&Field 10 & 64 & $7.6\times10^{10}$ & \multirow{3}{*}{0.53} & \multirow{3}{*}{H, N, O, Mg, Fe}  \\
&Field 11 & 64 & $7.6\times10^{10}$ &  & \\
&Field 12 & 64 & $5.3\times10^{10}$ &  &  \\
\hline
\multirow{4}{*}{06}
&Field 13 & 64 & $6.4\times10^{10}$ & \multirow{4}{*}{1.82} & \multirow{4}{*}{H, O, Mg, Fe} \\
&Field 14 & 64 & $6.4\times10^{10}$ &  &  \\
&Field 15 & 64 & $5.5\times10^{10}$ &  &  \\
&Field 16 & 64 & $1.4\times10^{11}$ &  &  \\
\hline
\multirow{5}{*}{07}
&Field 17 & 64 & $8.0\times10^{10}$ & \multirow{5}{*}{2.15} & \multirow{5}{*}{H, O, Mg, Fe}  \\
&Field 18 & 64 & $1.0\times10^{11}$ &  &  \\
&Field 19 & 64 & $4.4\times10^{10}$ &  & \\
&Field 20 & 64 & $3.0\times10^{11}$ &  &  \\
&Field 21 & 64 & $3.0\times10^{11}$ &  &  \\
\hline
\multirow{3}{*}{08}
&Field 22 & 64 & $2.1\times10^{11}$ & \multirow{3}{*}{3.04} & \multirow{3}{*}{H, O, Mg, Fe} \\
&Field 23 & 64 & $1.5\times10^{11}$ &  & \\
&Field 24 & 64 & $1.2\times10^{11}$ &  & 
\enddata
\tablecomments{The volume and most massive halo mass refer to the high-resolution region at $z=0$ in the corresponding DM-only realization.}
\end{deluxetable}

This appendix summarizes the individual zoom-in regions included in the current \darwinii\ suite. Table~\ref{tab:sim_list} lists the target region, the comoving volume of the high-resolution region at $z=0$, the mass of the most massive halo within that region, the latest redshift reached by each hydrodynamical simulation, and the chemical elements explicitly tracked. The volume and halo mass are measured at $z=0$ in the corresponding DM-only realization. Multiple field regions may be included in a single simulation run for computational efficiency.


\section{Optimization for OpenMP}\label{sub:omp} 


To fully utilize high-performance computing resources based on many-core architectures, we implement Open Multi-Processing (OMP) parallelism in \ramses\ in addition to the Message Passing Interface (MPI) layer. The OpenMP layer is independent of the MPI domain decomposition, provided that routines are thread-safe. To ensure thread safety, we remove the \texttt{save} attribute from local variables in the original \ramses\ code wherever possible, even at the cost of a minor performance penalty in some routines. Sequential \texttt{do}-loops are restructured to eliminate unnecessary serial operations, such as \texttt{atomic} directives, thereby distributing the  workload across multiple threads. The BH merging routine in \ramses\ is replaced by a tree-search-based implementation that is both thread-safe and parallel. In addition, we replace stack-allocated work arrays with dynamically allocated heap memory to avoid stack overflows caused by the large memory requirements of the \darwin\ runs.

The hybrid MPI--OpenMP parallelization is also implemented throughout the sub-grid modules introduced in Section~\ref{sub:subgrid}. Specifically, OpenMP parallelism is used for (i) the non-equilibrium primordial chemistry network coupled to three ionizing radiation bins and the M1-closure radiative transfer solver in \darwinii\ (Section~\ref{sub:cooling}), (ii) the local thermo-turbulent star formation prescription, for which the per-cell evaluation of the Mach number, virial parameter, and free-fall efficiency $\epsilon_{\rm ff}$ (Equation~\ref{eqn:eff}) is thread-parallelized over leaf cells, (iii) the mechanical Type II SN feedback scheme, with the metallicity- and density-dependent momentum injection computed independently for each host cell, (iv) the chemical enrichment routine that distributes ejecta from Type Ia and Type II SNe and AGB stars to the surrounding gas, and (v) the extended BH dynamical-friction module, including the contributions from gas, stars, and dark matter particles, as well as the dual jet–heating AGN feedback (Section~\ref{sub:agn}) and the associated spin evolution. Thread-level parallelization is not used for simple retrievals of pre-tabulated quantities, such as the photon production rates, the chemical yields, and {\sc cloudy} cooling and heating rates, because their computational cost is negligible relative to the physical update steps. OpenMP is therefore focused on the computationally intensive components of star formation, feedback, chemistry, radiative transfer, and BH dynamics, where the per-cell or per-particle operations dominate the runtime.

\begin{figure}
\vspace{2mm}
\centering 

\includegraphics[width=0.9\linewidth]{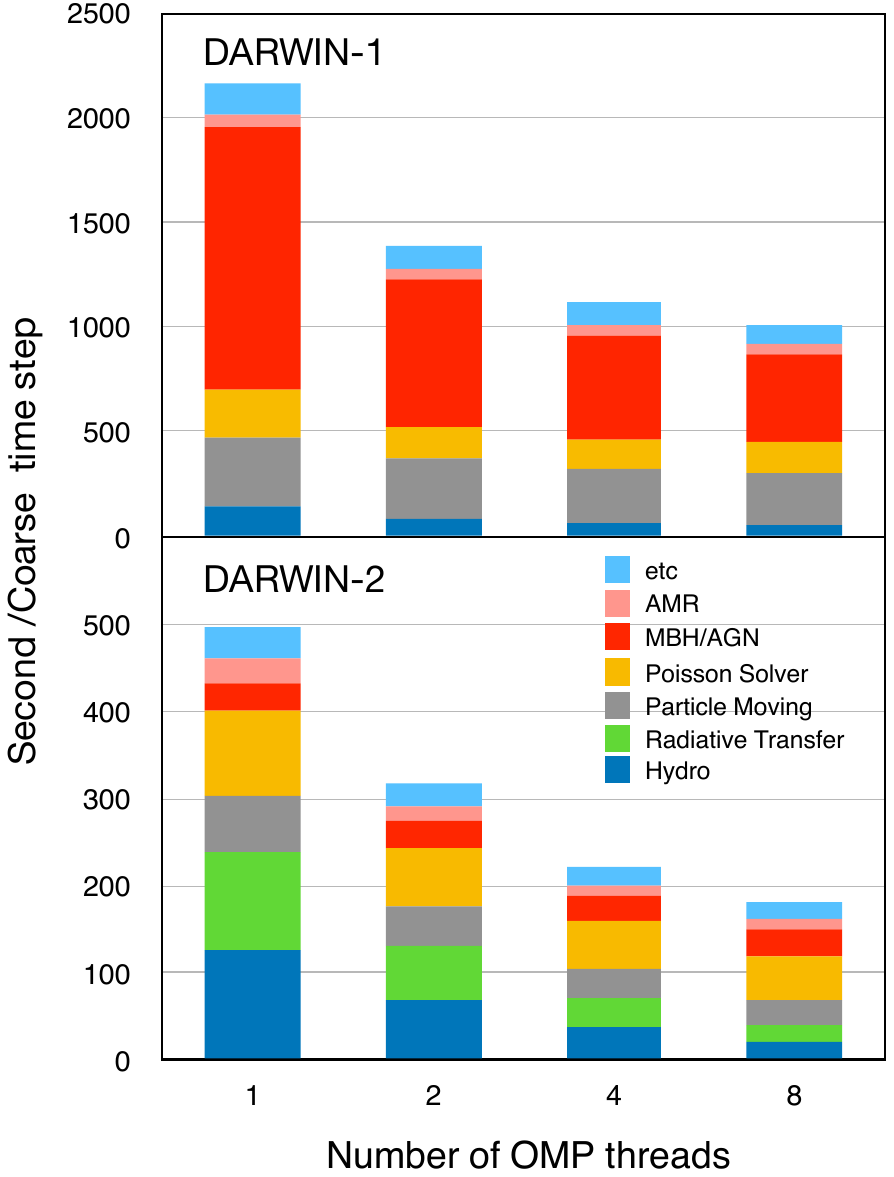}
\caption{OpenMP scaling of \darwini\ and one of the \darwinii\ runs. The \texttt{Hydro} and \texttt{Radiative Transfer} routines show good scaling in both simulations, while the \texttt{Particle Moving} and \texttt{AMR} routines exhibit more modest improvements. The \texttt{Poisson Solver} shows little to no scaling. The scaling behavior of the \texttt{MBH/AGN} routine differs between \darwinii\ and \darwini\: although the per-particle cost is higher in \darwinii\ due to the extended dynamical-friction treatment, the much smaller number of massive BH particles makes its fractional contribution to the total runtime subdominant.}

\label{fig:omp_test}

\end{figure}

In addition to the sub-grid modules, we revise the underlying \ramses\ infrastructure to improve parallelism in routines that were previously serial or only weakly threaded. The AMR tree traversal and cell-flagging procedures,  particle-to-mesh assignment, neighbor-search routines used by the mechanical-feedback and dynamical-friction modules, and the I/O operations for snapshot and light-cone dumps are restructured to reduce critical sections and enable fine-grained loop parallelism. The per-MPI-task memory footprint is further reduced by replacing static work arrays with dynamically allocated heap buffers, which is particularly important for \darwinii\ owing to the combined memory demands of radiative transfer, non-equilibrium chemistry, and the additional refinement levels (up to level 20).

The combined effect of these module-level and code-structure optimizations on the overall runtime is summarized in Figure~\ref{fig:omp_test}, which shows the scaling performance of the OpenMP-optimized code for two representative \darwin\ runs. In \darwini, the \texttt{MBH/AGN} routine dominates the total computing time when a single OpenMP thread is used. The \texttt{Hydro} routine scales particularly well in the Godunov solver because its strong spatial locality enables efficient thread-level parallelization of the flux calculations. Its overall performance, however, is limited by virtual boundary operations; because the Hilbert-curve-based domain decomposition requires all-to-all communication to synchronize ghost cells across MPI tasks, this communication overhead becomes increasingly important as the thread count grows. While \texttt{AMR}, \texttt{MBH/AGN}, and \texttt{Particle Moving} show moderate scaling, the computing time of the \texttt{Poisson Solver} routine decreases only modestly.

In \darwinii, \texttt{Hydro} and \texttt{Radiative Transfer} routines account for approximately half of the elapsed time per coarse time step when using a single OpenMP thread per MPI task. Consistent with the local nature of the Godunov-based hydrodynamics, the \texttt{Radiative Transfer} module also scales efficiently because the M1 closure primarily involves local flux exchanges between neighboring cells, reducing inter-process dependencies during physical updates. While \texttt{Particle Moving} and \texttt{AMR} show moderate scaling, the \texttt{MBH/AGN} routine contributes minimally to the total runtime and shows little improvement with increasing thread counts because \darwinii\ contains  relatively few massive BH particles. These contrasting scaling behaviors between \darwini\ and \darwinii\ indicate that the specific simulation setup and the balance between local computation and global communication significantly influence the overall efficiency of OpenMP parallelization.

\section{Merger tree}
\label{sec:appendix-tree}

\begin{figure*}
\centering 
\includegraphics[width=\linewidth]{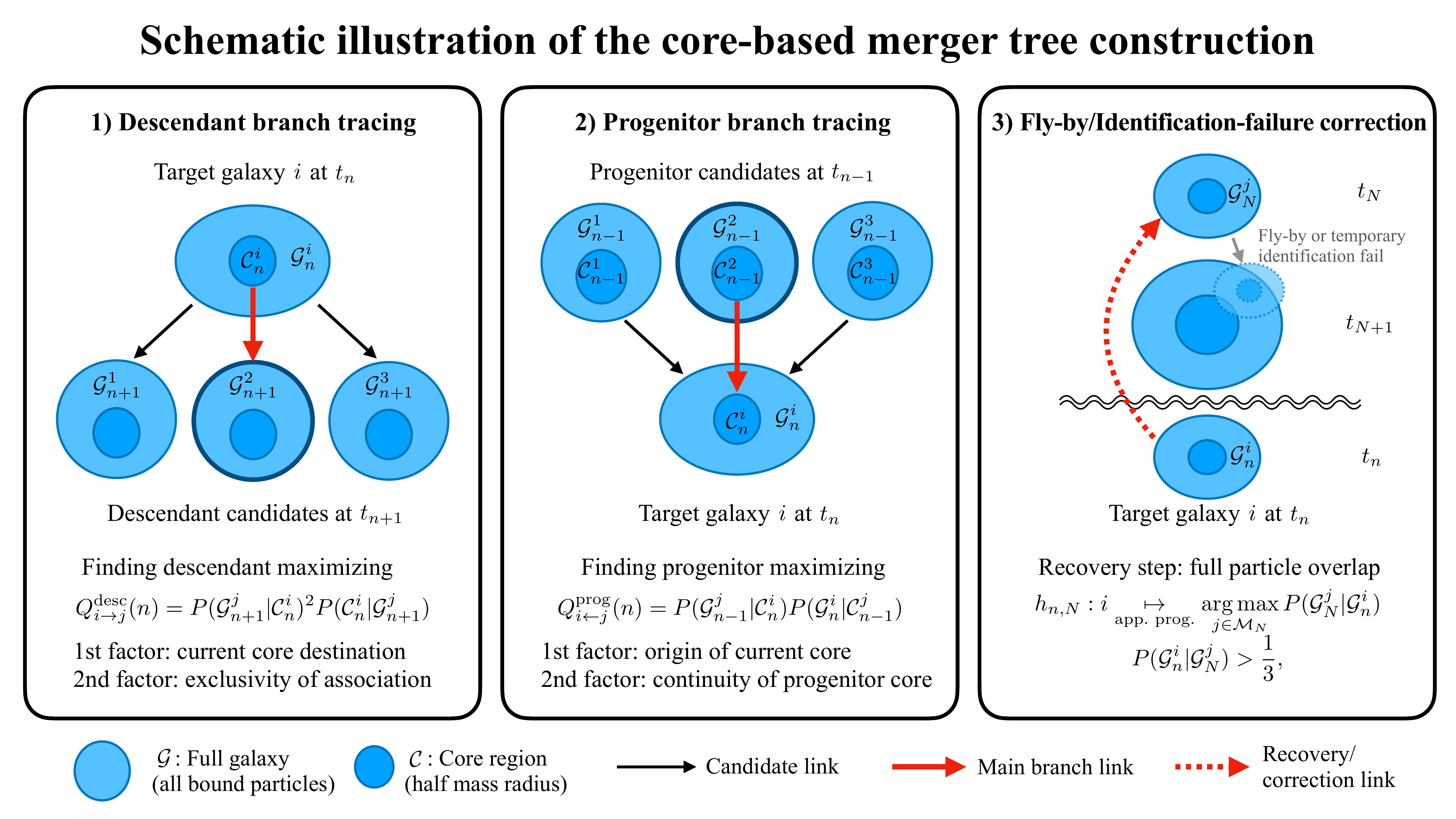}
\caption{Schematic illustration of the core-based merger tree construction in the \darwin\ simulations. The main branch connecting a target galaxy to its descendant or progenitor is identified by searching for the candidate maximizing the matching score evaluated from core and full stellar particles in galaxies. The branch correction for fly-by encounters or identification failures employs a full particle-overlap scheme to recover an appropriate progenitor.
} 
\label{fig:tree_scheme}
\end{figure*}

We trace the progenitors and descendants of galaxies and subhalos between adjacent snapshots using particle IDs. For galaxies, we use stellar particles, while for subhalos we use DM particles. In this section, we describe the branch-tracing scheme for galaxies, but the same procedure is applied to subhalos based on DM particles.

We formulate the branch-tracing procedure in terms of particle sets. 
%
We define $\mathcal{G}_n^i$ as the set of stellar particles assigned to the
$i^{\rm th}$ galaxy at time step $n$. For any set of particles $\mathcal{A}$, we denote the total mass of the particles in $\mathcal{A}$ evaluated at time step $n$ by $M_n(\mathcal{A})$. Since stellar particle masses evolve over time due to stellar mass loss, $M_n(\mathcal{A})$ can vary between time steps even for the same set of particle IDs. \texttt{PGalF} assigns each stellar particle to at most one galaxy
. 

For tree construction, we do not use all stellar particles assigned to the galaxy whose branch is being traced. Instead, for the $i^{\rm th}$ galaxy at time step $n$, we define a core particle set $\mathcal{C}_n^i\subset\mathcal{G}_n^i$ as
\begin{equation}
    \mathcal{C}_n^i= \left\{s\in\mathcal{G}_n^i\, \middle|\,|\mathbf{x}_s-\mathbf{x}_{\rm peak}^i|\le R_{\rm core}^i \right\},
\end{equation}
where $\mathbf{x}_{\rm peak}^i$ is the stellar density peak of the galaxy and
$R_{\rm core}^i$ is chosen so that $\mathcal{C}_n^i$ contains half the
stellar mass of $\mathcal{G}_n^i$. This choice follows the central stellar component of a galaxy and reduces spurious associations caused by particles in galaxy outskirts, where particle assignment can be affected by close pairs, fly-by, tidal features, or limitations of the galaxy finder. 

Figure~\ref{fig:tree_scheme} illustrates the branch-tracing and correction schemes in the \darwin\ simulations. In our scheme, the main progenitor or descendant is determined by using a core-based matching score defined as the product of two mass fractions. This criterion favors candidate associations that are both complete and exclusive, thereby suppressing spurious links to massive galaxies that contain only a small fraction of the target core.

For the $i^{\rm th}$ galaxy at time step $n$, we define the matching score with a descendant candidate $j$ at time step $n+1$ as
\begin{equation}
\label{eq:descen}
    Q_{i\rightarrow j}^{\rm desc}(n)=
    P(\mathcal{G}_{n+1}^j|\mathcal{C}_{n}^i)^{\alpha} 
    P(\mathcal{C}_{n}^i|\mathcal{G}_{n+1}^j).
\end{equation}
Here, $P$ denotes the mass-overlap fraction between two particle sets. For particle sets $\mathcal{A}_m$ and $\mathcal{B}_n$ identified at time steps $m$ and $n$, respectively, we define
\begin{equation}
P(\mathcal{A}_m|\mathcal{B}_n)
\equiv
\frac{M_n(\mathcal{A}_m\cap\mathcal{B}_n)}
{M_n(\mathcal{B}_n)}.
\end{equation}
The intersection is determined using particle IDs, and the particle masses are evaluated at the time step of the conditioning set, $\mathcal{B}_n$. Thus, $P(\mathcal{A}_m|\mathcal{B}_n)$ represents the fraction of the mass of $\mathcal{B}_n$ associated with particles that also belong to $\mathcal{A}_m$.
In Equation~\ref{eq:descen}, the first factor $P(\mathcal{G}_{n+1}^j | \mathcal{C}_n^i)$ measures how completely the core of galaxy $i$ is carried into a descendant candidate, whereas the second factor $P(\mathcal{C}_{n}^i | \mathcal{G}_{n+1}^j)$ measures how exclusively that candidate is associated with the core of galaxy $i$. 

In the descendant search, we weight the first factor more strongly ($\alpha = 2$ in Equation~\ref{eq:descen}) because the main descendant should represent the primary destination of the target core. We find that using the simple product, i.e., $\alpha = 1$, 
can occasionally favor a low-mass candidate that contains only a small fraction of the target core but is almost entirely composed of it, i.e., $P(\mathcal{G}_{n+1}^j | \mathcal{C}_n^i) \ll 1$ and $ P(\mathcal{C}_n^i | \mathcal{G}_{n+1}^j) \sim 1$. On the other hand, our 
empirical weighting reduces such misidentifications while preserving the overall continuity of the branches. This is particularly effective for subhalo merger trees, because the dark matter distributions of subhalos are more spatially extended and their identification is more susceptible to artificial fragmentation. Maximizing the product selects a descendant that contains most of the target core while avoiding spurious links to much larger systems that include only a small fraction of that core. 

The main descendant of galaxy $i$ is then defined as
\begin{equation}
    f_n(i)=
    \underset{j}{\arg\max}\,
    Q_{i\rightarrow j}^{\rm desc}(n).
\end{equation}
With this definition, multiple galaxies at time step $n$ may share a common
main descendant at time step $n+1$, i.e., $f_n(i)=f_n(k)$ for $i\ne k$, corresponding to merger events.

We identify the main progenitor in the reverse time direction using a related matching score. For the $i^{\rm th}$ galaxy at time step $n$ and a progenitor candidate $j$ at time step $n-1$, we define
\begin{equation}
\label{eq:prog}
    Q_{i\leftarrow j}^{\rm prog}(n)=
    P(\mathcal{G}_{n-1}^j|\mathcal{C}_{n}^i)
    P(\mathcal{G}_{n}^i|\mathcal{C}_{n-1}^j),
\end{equation}
where $P(\mathcal{G}_{n-1}^j|\mathcal{C}_{n}^i)$
is the fraction of the core stellar mass of galaxy $i$ at $t_n$ that was
contained in progenitor candidate $j$ at $t_{n-1}$, and $P(\mathcal{G}_{n}^i|\mathcal{C}_{n-1}^j)$
is the fraction of the core stellar mass of candidate $j$ at $t_{n-1}$ that
contributes to galaxy $i$ at $t_n$. 
The tentative main progenitor is
\begin{equation}
\tilde{g}_n(i)=
\underset{j}{\arg\max}\,
Q_{i\leftarrow j}^{\rm prog}(n).
\end{equation}
We accept $\tilde{g}_n(i)$ as the main progenitor, $g_n(i)$, only if its main descendant is the target galaxy itself,
\begin{equation}
f_{n-1}\left[\tilde{g}_n(i)\right]=i.
\end{equation}
This consistency condition also ensures that two galaxies at the same time step cannot share the same main progenitor, thereby preventing a single progenitor from fragmenting into multiple main branches.

\begin{figure}
\centering 
\includegraphics[width=\linewidth]{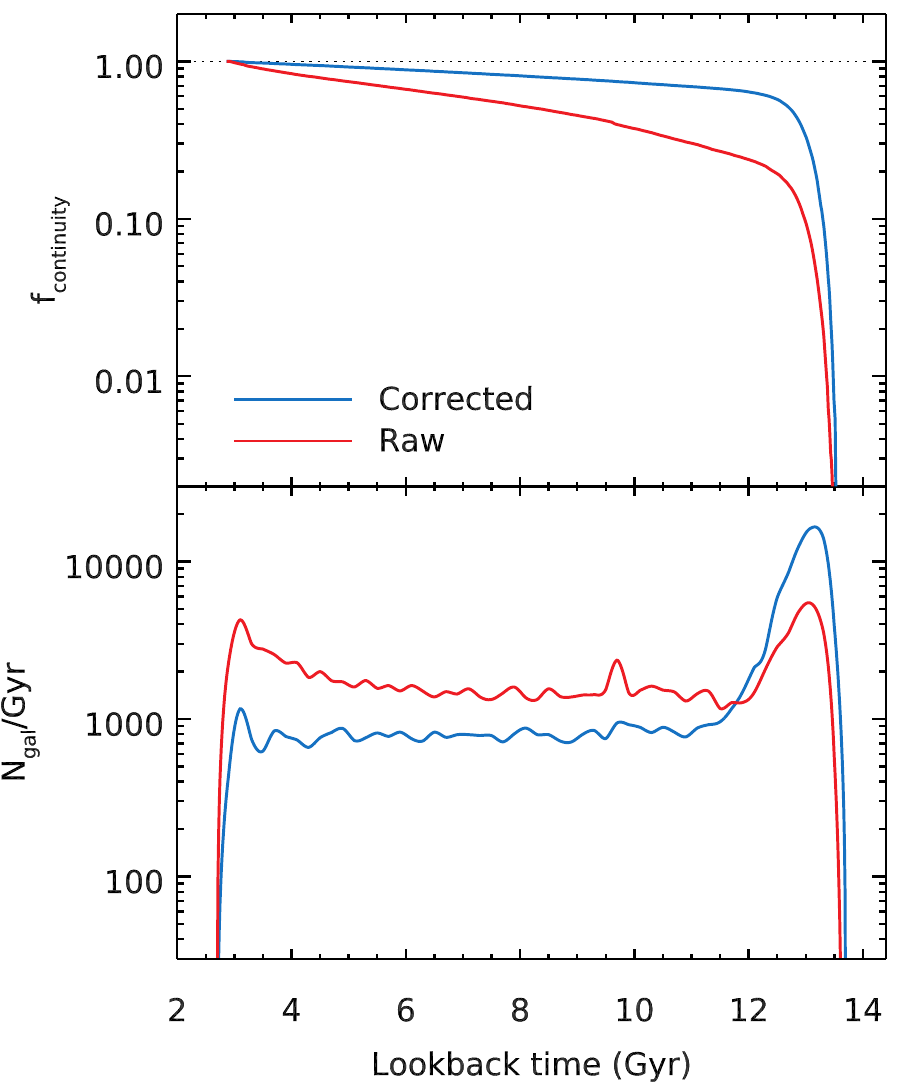}
\caption{{\it Top:} Continuity of the corrected (blue) and raw (red) merger trees for galaxies with $M_\star>10^8\,\msun$ in the final snapshot of \darwini\ as a function of lookback time. {\it Bottom:} 
Number of branches first identified per unit lookback time. The corrected trees allow galaxy branches to be traced back to earlier cosmic times more robustly.}
\label{fig:tree_comp}
\end{figure}

By contrast, the reverse condition, $g_{n+1}[f_n(i)]=i$, is not generally required. While a single descendant galaxy may have multiple progenitors during a merger, only the primary ancestor is designated as the main progenitor. The consistency condition may fail, however, when a galaxy undergoes a close fly-by or is temporarily missed by the structure finder. In such cases, the adjacent-snapshot search cannot identify a consistent main progenitor, and we apply the branch-correction procedure described below.

To identify the appropriate progenitors of such fly-by or temporarily missing galaxies, we revisit earlier time steps with $N\le n-2$. In this correction step, we do not use the core particle sets or the core-based matching scores defined above. Instead, we use the full stellar particle set of the galaxy at $t_n$ and search among galaxies at $t_N$ whose branches merge into or disappear within another system at $t_{N+1}$. Denoting this candidate set by $\mathcal{M}_N$, we define
\begin{equation}
h_{n,N}(i)=
\underset{j\in\mathcal{M}_N}{\arg\max}\,
P(\mathcal{G}_N^j|\mathcal{G}_n^i).
\end{equation}
Here, $P(\mathcal{G}_N^j|\mathcal{G}_n^i)$ measures the fraction of the stellar mass of galaxy $i$ at $t_n$ that was already contained in candidate $j$ at $t_N$.

A nonzero overlap can occasionally arise from a relatively small subset of particles exchanged between galaxies. To ensure that the earlier candidate itself is substantially associated with the target galaxy, we additionally require
\begin{equation}
P(\mathcal{G}_n^i|\mathcal{G}_N^j)>\frac{1}{3}.
\end{equation}
This condition requires that more than one-third of the stellar mass of the earlier candidate is subsequently found in the target galaxy.

We apply this procedure over a lookback time interval satisfying $t_n-t_N<1\,$Gyr, proceeding from earlier to later time steps. Once an appropriate progenitor is identified for a fly-by or temporarily unidentified system, we reconnect the branch by adding a recovery link between the identified progenitor at $t_N$ and the target galaxy at $t_n$, thereby bridging the temporarily missing interval. 
Figure~\ref{fig:tree_comp} demonstrates the performance of the branch-correction procedure. We follow the main progenitor branches of galaxies with $M_\star>10^8\,\msun$ in the final snapshot of \darwini\ backward in time. In the top panel, $f_{\rm continuity}$ is defined as the fraction of these final-snapshot galaxies whose branches remain connected at a given lookback time. Compared with the raw trees, the corrected trees maintain branch continuity to significantly earlier cosmic times. The lower panel shows the number of branches first identified when traced backward in time, indicating that the correction procedure substantially reduces artificial branch truncation caused by fly-by events or temporary identification failures. Nevertheless, some galaxies still contain old stellar particles whose progenitor branches remain unrecovered after the branch-correction procedure. These cases are mostly associated with objects that were not identified by the structure-finding algorithm in earlier snapshots, indicating that the recovery of disrupted branches is ultimately limited by the completeness of the underlying structure catalog.


\section{Machine-learning Methodology}
\label{sec:appendix-ml}

\begin{figure}
  \centering
  \includegraphics[width=\linewidth]{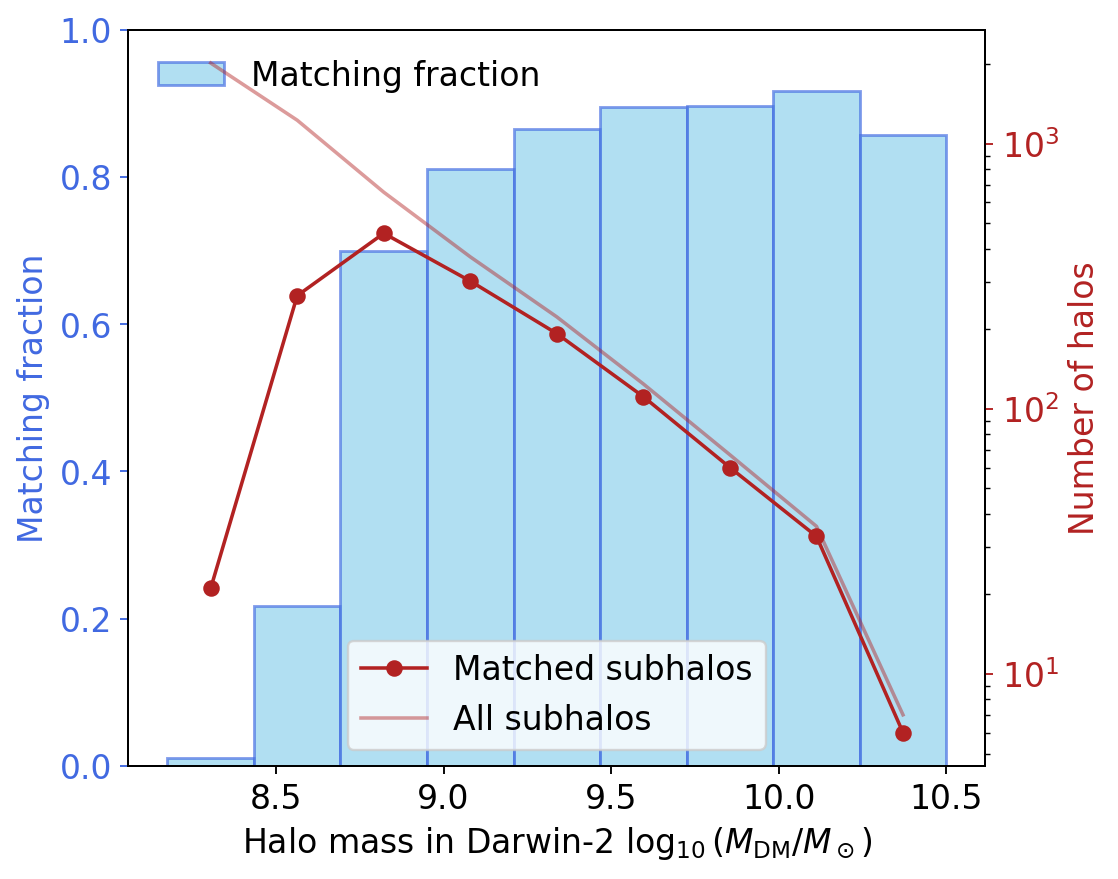}
  \caption{The number of \darwinii\ halos as a function of dark matter mass, along with the subset that is bijectively matched (solid lines). The matching fraction in each mass bin is shown as a histogram.}
  \label{fig:matching_frac}
\end{figure}

\subsection{Particle-based Halo Matching}

This section summarizes the construction of the matched \darwini--\darwinii\ catalog and the ML pipeline used in the proof-of-concept resolution-bias study presented in
Section~\ref{sub:ML}, focusing on the changes relative to \cite{2024ApJ...965..156J}. For the main benchmark, we use the \darwini\ snapshot and eight \darwinii\ zoom fields near $z=3.86$--$3.89$. The final benchmark sample contains 1454 bijective pairs across these eight fields, of which 302 have $M_{\star,\rm D2}>0$.

The starting point of the ML analysis is a halo-by-halo matched catalog between \darwini\ and \darwinii. The matching itself is based on dark-matter particle membership. For each halo we first identify the central DM particles within its dark matter halo. Because \darwini\ is three refinement levels coarser than \darwinii, we use the 20 most central dark-matter particles in \darwini\ and the 1000 most central dark-matter particles in \darwinii. We find the nearest dark-matter particle in the IC distribution. A forward match is accepted only if the target halo contains at least half of the mapped central particles. We repeat the same procedure in the reverse direction and retain only reciprocal matches as the final bijective \darwini--\darwinii\ pairs. 

Figure \ref{fig:matching_frac} displays the total number of halos across all \darwinii\ fields at $z=3.9$, including those with $M_{\star}=0$. More than 85\% of galaxies with $M_{\rm DM} > 10^9 
\,\mathrm{M}_{\odot}$ are bijectively matched.

\subsection{Machine-learning Model and Training}

For the machine-learning model, we use 11 features, including 7 features directly derived from the dark matter and baryonic content of the galaxy: the total masses of dark matter, stars, gas, cold ($T<10^{4}\,{\rm K}$) gas, and black holes, as well as the dark matter half-mass radius and gas-phase metallicity. In addition, four environmental features are adopted from \cite{2019MNRAS.489.3565J}: local density, number of nearby halos, sum of mass over distance, and maximum mass over distance. We use \texttt{GroupKFold}, with six fields for training, one for validation, and one for testing. Early stopping is applied using the validation set to identify the optimal number of iterations. 
Incorporating the validation field, the model is then retrained using the optimal iteration threshold established in the previous step.

\bibliographystyle{aasjournal}
\bibliography{references}

\end{document}